\documentclass[12pt]{iopart}

 \newcommand\varpm{\mathbin{\vcenter{\hbox{%
  \oequation{\hfil$\scriptstyle({+})$\hfil\cr
          \noequation{\kern-.3ex}
          \hfil$\scriptscriptstyle{-}$\hfil\cr}%
}}}}

\usepackage{cite}
\usepackage{graphicx}
\begin{document}

\title{Pair-excitation energetics of highly correlated many-body states}


\author{M. Mootz, M. Kira, and S.W. Koch}
\address{Department of Physics and Material Sciences Center, Philipps-University Marburg,
Renthof 5, D-35032 Marburg, Germany}

\ead{martin.mootz@physik.uni-marburg.de}

\begin{abstract}
A microscopic approach is developed to determine the excitation energetics of highly correlated
quasi-particles in optically excited semiconductors based entirely on a pair-correlation function
input. For this purpose, the Wannier equation is generalized to compute the energy per excited
electron--hole pair of a many-body state probed by a weak pair excitation. The scheme is verified
for the degenerate Fermi gas and incoherent excitons. In a certain range of experimentally
accessible parameters, a new stable quasi-particle state is predicted which consists of four to six
electron--hole pairs forming a liquid droplet of fixed radius. The energetics and
pair-correlation features of these ''quantum droplets'' are analyzed.   
\end{abstract}

\pacs{73.21.Fg,71.10.-w,71.35.-y}
\maketitle

\newcommand{\be}{\begin{equation}}
\newcommand{\ee}{\end{equation}}
\newcommand{\bea}{\begin{eqnarray}}
\newcommand{\eea}{\end{eqnarray}}
\newcommand{\eps}{\varepsilon}
\newcommand{\ev}[1]{\langle#1\rangle}
\newcommand{\ddt}{\frac{\partial}{\partial t}}
\newcommand{\ihddt}{i\hbar\frac{\partial}{\partial t}}
\newcommand{\mcal}[1]{{\mathcal{#1}}}
\newcommand{\mrm}[1]{{\mathrm{#1}}}
\newcommand{\drm}{{\mathrm{d}}}
\renewcommand{\matrix}[1]{\mathbf{#1}}
\renewcommand{\vec}[1]{\mathbf{#1}}
\newcommand{\ci}[1]{\mathbf{#1}}
\newcommand{\stumm}{\bullet}
\newcommand{\ssum}[1]{\left[\left[#1\right]\right]}
\newcommand{\csum}[2]{\left|\left|#2\right|\right|_{#1}}
\newcommand{\widebar}[1]{\overline{#1}}
\newcommand{\re}{\mathrm{Re}}
\newcommand{\im}{\mathrm{Im}}
\newcommand{\op}[1]{\hat{#1}}
\newcommand{\ave}[1]{\langle #1 \rangle}
\newcommand{\comm}[2]{\left[ #1,\,#2 \right]_{-}}
\newcommand{\trace}[1]{{\rm Tr}\left[ #1 \right]}

\newcommand{\absN}[1]{{| #1 \rangle}}
\newcommand{\absNS}[1]{{\langle #1 |}}
\newcommand{\absPROJ}[2]{{\langle {#1} | {#2} \rangle}}

\newcommand{\ident}[0]{\mathbb{I}}


\section{Introduction}

Interactions may bind matter excitations into new stable entities,
quasi-particles, that typically have very different properties than the noninteracting constituents. In
semiconductors, electrons in the conduction band and vacancies, i.e.~holes, in the valence band
attract each other via the Coulomb interaction \cite{Book:11}. Therefore, the Coulomb attraction may
bind different numbers of electron--hole pairs into a multitude of quasi-particle configurations.
The simplest example is an exciton \cite{Frenkel:31,Wannier:37} which consists of a
Coulomb-bound electron--hole pair and exhibits many analogies to the hydrogen atom \cite{Book:11}. Two excitons can 
bind to a molecular state known as the biexciton \cite{Miller:82,Kim:94}. Both, exciton and
biexciton resonances can be routinely accessed in present-day experiments by exciting a high quality
direct-gap semiconductor
optically from its ground state.  Even the exciton
formation can directly be observed in both optical \cite{Khitrova:99} and terahertz (THz)
\cite{Kaindl:03} spectroscopy and their abundance can be controlled via the intensity of the optical
excitation \cite{Smith:10}. Also higher correlated quasi-particles can emerge in semiconductors. For
instance, polyexcitons or macroscopic electron--hole droplets have been detected
\cite{Steele:87,Turner:10,Jeffries:75,Wolfe:75}, especially in semiconductors with an indirect gap.

To determine the energetics of a given quasi-particle configuration, one can apply density-functional theory based on the functional dependence of the total energy on
the electron density \cite{Fiolhais:03,Sholl:09}. This procedure is well established in particular
for ground-state properties. However, whenever one wants to model experimental signatures of excited
quasi-particle states in the
excitation spectra, the applicability of density-functional theory becomes challenging,
especially for highly correlated states. 

\begin{figure}[t]
\includegraphics*[scale=0.65]{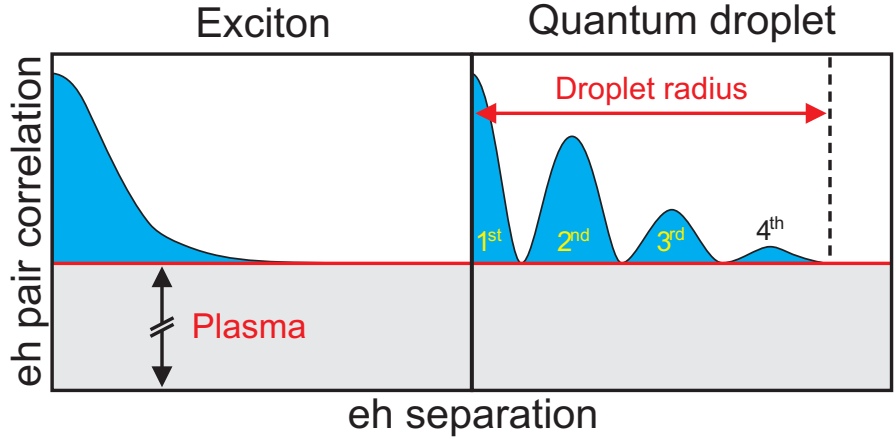}
\centering
\caption{Schematic representation of the exciton (left) and the quantum droplet electron--hole
(eh) pair-correlation function $g(\vec{r})$. The plasma contribution (gray shaded area) is shown
together  with the correlation contribution (blue shaded area). The radius of the quantum droplet is
indicated by the vertical dashed line and each of the rings are labeled.} 
\label{Fig1}
\end{figure}

In this paper, we develop a new scheme
to determine the excitation energetics of highly correlated quasi-particle configurations. We start
directly from the pair-correlation function, not from the density functional,
and formulate a framework to compute the pair-excitation energetics. The electron--hole
pair-correlation function $g(\vec{r})$ defines the conditional probability of finding an
electron at the position $\vec{r}$ when the hole is at the origin. As an example, we show in  
figure \ref{Fig1} examples of $g(\vec{r})$ for excitons (left) and quantum droplets (right). Here,
we refer to quantum droplets as a quasi-particle state where few electron--hole pairs, typically
four to six, are in a liquid-like state bounded within a sphere of microscopic radius $R$. 

In general, $g(\vec{r})$
always contains a constant electron--hole plasma contribution (gray shaded area) stemming from the
mean-field aspects of the many-body states. The actual bound quasi-particles are described by the
correlated part $\Delta g(\vec{r})$ (blue shaded area) which decays for increasing electron--hole
separation. For $1s$ excitons, $\Delta g(\vec{r}) \propto |\phi_{1s}(\vec{r})|^2$ decreases
monotonically and has the shape defined by the $1s$-exciton wave function $\phi_{1s}(\vec{r})$
\cite{Kira:06b}. Since the electrons and holes in a quantum droplet are in a liquid phase, $\Delta
g(\vec{r})$ must have the usual liquid structure where particles form a multi-ring-like pattern
where the separation between the rings is defined by the average particle distance
\cite{narten:72,Jorgensen:83,Fois:94}. Due to the electron--hole attraction, one also observes a
central peak, unlike for single-component liquids. 

We derive the pair-excitation energetics
for an arbitrary initial many-body state in section~\ref{Sec:approach}. In this connection,
we first study the pair excitations of the semiconductor ground state before we extend the
approach for an arbitrary initial many-body state. We then test our approach for the well-known
cases of a degenerate Fermi gas and incoherent excitons in section~\ref{Sec:calc}. In
section~\ref{Sec:Epair_drop}, we apply our scheme to study the energetics and structure of quantum
droplets based on electron--hole correlations in a GaAs-type quantum well (QW). The effect of
carrier--carrier correlations on the quantum droplet energetics is analyzed in
section~\ref{Sec:ee_corr}.

\section{Energy and correlations in many-body systems}
\label{Sec:approach}

For resonant excitations, the excitation properties of many direct-gap semiconductor QW systems can
be modeled using a two-band Hamiltonian \cite{Kira:99,Book:11}
\begin{equation}
\label{eq:system_Ham}
\hat{H}=\sum_{\vec{k},\lambda}\epsilon_{\vec{k}}^{\lambda}a^{\dagger}_{\lambda,\vec{
k}}a_{\lambda,\vec{k}}+\frac{1}{2}\sum_{\vec{k},\vec{k}',\vec{q},\lambda,\lambda'}V_{\vec{q}}\,
a^{\dagger}_{\lambda,\vec{k}+\vec{q}}a^{\dagger}_{\lambda',\vec{k}'-\vec{q}}a_{\lambda',\vec{k}'}a_{
\lambda,\vec{k}}\,.
\end{equation}
where the Fermionic operators $a^{\dagger}_{v(c),\vec{k}}$ and $a_{v(c),\vec{k}}$ create
and annihilate an electron with crystal momentum $\hbar\,\vec{k}$ in the valence (conduction) band,
respectively. We consider excitations close to the $\Gamma$ point such that the kinetic energies can be treated as parabolic
\begin{equation}
 \epsilon_{\vec{k}}^{c}=\frac{\hbar^2\vec{k}^2}{2m_e}+\mathrm{E}_\mathrm{g}\,,\qquad
\epsilon_{\vec{k}}^{v}=-\frac{\hbar^2\vec{k}^2}{2m_h}\,,
\end{equation}
with the bandgap energy $E_{\mrm{g}}$ and the effective masses for the
electron $m_{e}$ and hole $m_{h}$. The Coulomb
interaction is characterized by the matrix element $V_{\vec{q}}$ of the quantum confined system
\cite{Book:11}. We have formally set $V_{\vec{q}=0}=0$ to eliminate the $\vec{q}=0$ contribution
from the Coulomb sum, which enforces the overall charge neutrality in the system\cite{Book:11}.

For later use, we introduce Fermion field operators without the lattice-periodic functions
\begin{equation}
\label{eq:field}
\hat{\Psi}_{e}(\vec{r})=\frac{1}{\sqrt{S}}\sum_{\vec{k}}a_{c,\vec{k}}\,\e^{\mrm{i}\vec{k}
\cdot\vec { r }
}\,,\qquad
\hat{\Psi}_{h}(\vec{r})=\frac{1}{\sqrt{S}}\sum_{\vec{k}}a^{\dagger}_{v,\vec{k}}\,\e^{
-\mrm { i }
\vec { k }
\cdot\vec { r }
}\,,
\end{equation}
for electrons and holes, respectively. These can be directly used to follow e.g. electron (hole)
densities $\rho_{e(h)}(\vec{r})\equiv\ev{\hat{\Psi}^{\dagger}_{e(h)}(\vec{r})\hat{\Psi}_
{e(h)}(\vec{r})}$ on macroscopic length scales because the unit-cell dependency is already
averaged over. The corresponding normalization area is given by $S$.

\subsection{Ground-state pair excitations}
\label{sec:method}

\begin{figure}[t]
\includegraphics*[scale=0.65]{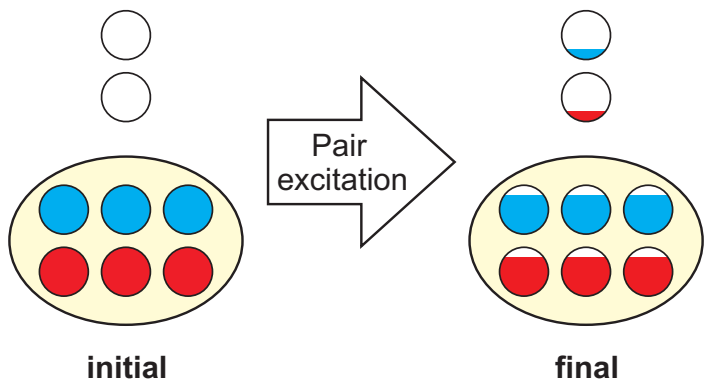}
\centering
\caption{Schematic representation of a pair excitation. The quasi-particle configuration is shown
before (left) and after (right) the pair excitation. Electron (holes) are symbolized by blue (red)
circles while a yellow ellipse surrounds the correlated pairs. The level of filling indicates the
fraction of electrons and holes bound as correlated pairs.}
\label{Fig_pe}
\end{figure}

A schematic representation of a pair excitation is shown in figure~\ref{Fig_pe} to illustrate the
detectable energetics. The individual electrons and holes are symbolized by circles while the
yellow ellipse surrounds the correlated pairs. The level of blue (red) filling indicates the
fraction of electrons (holes) bound as correlated pairs within the entire many-body system. This
fraction can be changed continuously by applying, e.g. an optical field to generate pair
excitations. If all pairs are bound to a single quasi-particle type, the initial energy of the
system is  
\begin{equation}
\label{eq:E_ini}
  E_{\mrm{ini}}=N\,E(N)\,,
\end{equation}
where $N$ is the total number of pairs. Since $N$ is typically much larger than the number of pairs
within a quasi-particle, it is meaningful to introduce $E(N)$ as the binding energy per excited
electron--hole pair. For stable quasi-particle configurations, a change in $N$ does not alter
$E(N)$, yielding the stability condition $\textstyle{\frac{\partial E(N)}{\partial N}}=0$.

We now assume that only a small number of pairs, $\delta N$, is excited from the quasi-particle
into an unbound pair. An example of the excited configuration is presented in the right panel of
figure~\ref{Fig_pe}. This state has the energy
\begin{eqnarray}
\label{eq:E_pair}
  E_{\mrm{fin}}&=(N-\delta N)\,E(N-\delta N)+\delta N E_{\mrm{pair}} \nonumber \\
   &=N E(N)+\delta N(E_{\mrm{pair}}-E(N))+\delta N\frac{\partial E(N)}{\partial
N}+\mathcal{O}(\delta N^2)\,,
\end{eqnarray}
where $E_{\mrm{pair}}$ is the energy of the unbound pair. After we apply the stability condition
$\textstyle{\frac{\partial E(N)}{\partial N}}=0$, we find that the pair excitation produces an
energy change $\Delta E\equiv E_{\mrm{fin}}-E_{\mrm{ini}}=\delta
N(E_{\mrm{pair}}-E(N))+\mathcal{O}(\delta N^2)$ such that the energy per excited particle becomes
\begin{eqnarray}
\label{eq:res}
  \bar{E}=\lim_{\delta N \to 0}\frac{\Delta E}{\delta N}=E_{\mrm{pair}}-E(N)\,.
\end{eqnarray}
This difference defines how much energy the electron--hole pair gains by forming the quasi-particle from
unbound pairs. 

To develop a systematic method describing the quasi-particle energetics, we start from the simplest
situation where the unexcited semiconductor is probed optically, i.e.~by inducing a weak pair
excitation. The corresponding
initial state is then the semiconductor's ground state $|G\rangle$ where all valence bands are fully
occupied while all conduction bands are empty. Following the analysis in
reference~\cite{Kira:06b}, we introduce the coherent displacement-operator functional
\cite{Book:11,Kira:06b}
\begin{equation}
\label{eq:disp}
  \hat{D}[\psi]=\e^{\varepsilon\hat{S}[\psi]}\,, \quad
\hat{S}[\psi]=\sum_{\vec{k}}\left(\psi_{\vec{k}}a^{\dagger}_{c,\vec{k}}a_{v,\vec{k}}
-\psi_ { \vec { k
} }^{\star}a^{\dagger}_{v, \vec{k}}a_{c,\vec{k}}\right)\,,
\end{equation}
to generate pair excitations. Here, $\varepsilon$ is an infinitesimal constant and $\psi_{\vec{k}}$ is a function to be
determined later using a variational approach. The probed ground state has a density matrix
$\hat{\rho}_{\mrm{G}}$ that determines the pair-excitation state via
\begin{equation}
\label{eq:rho}
  \hat{\rho}[\psi]=\hat{D}[\psi]\,\hat{\rho}_{\mathrm{G}}\, \hat{D}^{\dagger}[\psi]\,.
\end{equation}
We see from the definition (\ref{eq:disp}) that $\hat{D}[\psi]$
generates pair excitations to the semiconductor ground state $\hat{\rho}_{\mrm{G}}$ because
$\hat{S}[\psi]$ contains all elementary, direct, pair-excitation processes
$a^{\dagger}_{c,\vec{k}}a_{v,\vec{k}}$ ($a^{\dagger}_{v,\vec{k}}a_{c,\vec{k}}$) where an electron is
moved from the valence (conduction) to the conduction (valence) band. The weak excitation of the
probe is realized by making $\varepsilon$ infinitesimal, i.e. $\varepsilon \ll 1$.

As shown in reference~\cite{Kira:06b}, the pair excitation (\ref{eq:disp}) generates the
electron--hole
distribution and polarization
\begin{eqnarray}
\label{eq:probe_np}
f_{\vec{k},\psi}\equiv\mrm{Tr}\left[a^{\dagger}_{c,\vec{k}}a_{c,\vec{k}}\,
\hat{\rho}[\psi]\right]\equiv\mrm{Tr}\left[a_{v,\vec{k}}a^{\dagger}_{v,\vec{k}}\,
\hat{\rho}[\psi]\right]=\sin^2(\varepsilon|\psi_ { \vec { k } } |)\,, \nonumber \\
P_{\vec{k},\psi}\equiv\mrm{Tr}\left[a^{\dagger}_{v,\vec{k}}a_{c,\vec{k}}\,\hat{\rho}[\psi]\right]
=\e^ { \mrm { i } \varphi_ { \vec { k } } } \sin(\varepsilon\, |\psi_ { \vec { k } }|)
\cos(\varepsilon\,|\psi_{\vec{k}}|)\,,
\end{eqnarray}
respectively. Here, $\psi_{\vec{k}}=|\psi_{\vec{k}}|\e^{\mrm{i}\phi_{\vec{k}}}$ has been defined in
terms of a real-valued amplitude $|\psi_{\vec{k}}|$ and phase $\phi_{\vec{k}}$. For the
weak-excitation limit $\varepsilon\ll 1$, equation~(\ref{eq:probe_np}) reduces to
\begin{equation}
\label{eq:probe_np_weak} 
  f_{\vec{k},\psi}=\varepsilon^2|\psi_{\vec{k}}|^2+\mathcal{O}(\varepsilon^3)\,,\qquad
P_{\vec{k},\psi}=\varepsilon\, \psi_ { \vec { k } }+\mathcal{O}(\varepsilon^3)\,,
\end{equation}
to the leading order. Also the exact energy of state $\hat{\rho}[\psi]$ has already been computed in
reference~\cite{Kira:06b} with the result
\begin{eqnarray}
\label{eq:pair_exc} \fl
E_{\mrm{pro}}[\psi]\equiv
E[\psi]-E_{\mrm{GS}}=\mrm{Tr}\left[\hat{H}\hat{\rho}[\psi]\right]-\mrm{Tr} \left [
\hat{H}\hat{\rho}_{\mrm{G}}\right] \nonumber \\
\fl\qquad\quad\,\,=\varepsilon^2\left(\sum_{\vec{k}}\frac{\hbar^2\vec{k}^2}{2\mu}|\psi_{\vec{k}}
|^2-\sum_ { \vec
{k},\vec{k}'}V_{\vec{k}-\vec{k}'}\psi_{\vec{k}}\psi_{\vec{k}'}^{\star}\right)+\mathcal{O}
(\varepsilon^3)\,, \qquad \mu\equiv\frac{m_{e}m_{h}}{m_{e}+m_{h}}\,,
\end{eqnarray}
where we removed the ground-state energy $E_{\mrm{GS}}$ and introduced the reduced mass $\mu$.

\subsection{\label{sec:ord_Wannier} Ordinary Wannier equation}

The lowest pair-excitation energy can be found by minimizing $E_{\mrm{pro}}[\psi]$ with the
constraint that the number of excited electron--hole pairs 
\begin{equation}
\label{eq:Npro_G}
N_{\mrm{pro}}\equiv \sum_{\vec{k}}
f_{\vec{k},\psi}=\varepsilon^2\sum_{\vec{k}}|\psi_{\vec{k}}|^2
\end{equation}
remains constant. This can be accounted for by the standard procedure of introducing a Lagrange
multiplier $E_{\lambda}$ to the functional
\begin{equation}
 \label{eq:functional}
 F[\psi]\equiv
E_{\mrm{pro}}[\psi]-E_{\lambda}\varepsilon^2\sum_{\vec{k}}|\psi_{\vec{k}}|^2\,.
\end{equation}
By demanding $\delta F[\psi]=0$ under any infinitesimal change $\psi_{\vec{k}}\, \rightarrow
\, \psi_{\vec{k}}+\delta \psi_{\vec{k}}$, this extremum condition produces the Wannier
equation \cite{Kira:06b}
\begin{equation}
 \label{eq:Wannier}
\frac{\hbar^2\vec{k}^2}{2\mu}\psi_{\vec{k}}-\sum_{\vec{k}'}V_{\vec{k}-\vec{k}'}\psi_{\vec{k}'}=E_{
\lambda}\psi_{\vec{k}}\,.
\end{equation}
Fourier transform of equation~(\ref{eq:Wannier}) produces the real-space form
\begin{equation}
 \label{eq:Wannier_r}
\left[-\frac{\hbar^2\nabla^2}{2\mu}-V(\vec{r})\right]\psi(\vec{r})=E_{\lambda}\psi(\vec{r})\,,
\end{equation}
where $V(\vec{r})$ and $\psi(\vec{r})$ are the Fourier transformations of $V_{\vec{k}}$ and
$\psi_{\vec{k}}$, respectively. Since equations~(\ref{eq:Wannier}) and (\ref{eq:Wannier_r}) are the
usual Wannier equations for excitons, the exciton wave function defines those pair
excitations that produce minimal energy $E_{\lambda}$. At the same time,
equation~(\ref{eq:Wannier_r}) is
fully analogous to the Schr\"odinger equation of atomic hydrogen \cite{Book:11}. Therefore,
$E_{\lambda}$ also defines the Coulombic binding energy of excitons.

For the identification of the quasi-particle energy, we use the result (\ref{eq:res}) and compute the energy per excited electron--hole pair
\begin{equation}
 \label{eq:E_per}
\bar{E}_{\mrm{pro}}\equiv\frac{E_{\mrm{pro}}}{N_{\mrm{pro}}}\,.
\end{equation}
By inserting the solution (\ref{eq:Wannier}) into equations~(\ref{eq:pair_exc}) and
(\ref{eq:Npro_G}), we find $\bar{E}_{\mrm{pro}}=E_{\lambda}$ showing that the energetics of the pair-excitations from the ground state are defined by the exciton resonances. As a result, the energy per
probe-generated electron--hole pair produces a series of exciton resonances that can be detected,
e.g.~in the absorption spectrum. We will show next that this variational approach can be generalized
to determine the quasi-particle energetics for any desired many-body state.

\subsection{Average carrier-excitation energy}

Here, we start from a generic many-body system defined by the density matrix $\hat{\rho}_{\mathrm{MB}}$
instead of the semiconductor ground state $\hat{\rho}_{\mrm{G}}$. We assume that
$\hat{\rho}_{\mathrm{MB}}$ contains spatially homogeneous excitations with equal numbers of
electrons and holes, i.e. 
\begin{equation}
 \label{eq:Neh} \fl
 N_{eh}=\sum_{\vec{k}}f^{e}_{\vec{k}}=\sum_{\vec{k}}f^{h}_{\vec{k}}\,, \qquad
\mrm{with} \qquad f^{e}_{\vec{k}}\equiv
\ev{a^{\dagger}_{c,\vec{k}}a_{c,\vec{k}}}\,, \qquad f^{h}_{\vec{k}}\equiv
1-\ev{a^{\dagger}_{v,\vec{k}}a_{v,\vec{k}}}\,,
\end{equation}
where the electron (hole) distribution $f^{e}_{\vec{k}}$
($f^{h}_{\vec{k}}$) is defined within the electron--hole picture \cite{Book:11}. In general,
each electron--hole pair excitation increases 
the energy by $E_{\mrm{g}}$ because an electron is excited from the valence to the conduction band.
To directly monitor the energetics of $\hat{\rho}_{\mrm{MB}}$, we remove the
trivial $E_{\mrm{g}}N_{eh}$ contribution, yielding the average carrier energy
\begin{eqnarray}
\label{eq:Emb} \fl
 E_{\mrm{MB}}\equiv\ev{\hat{H}}-E_{\mrm{g}}N_{eh}=
\mathrm{Tr}\left[\hat{H}\,\hat{\rho}_{\mrm{MB}}\right]-E_{\mrm{g}}N_{eh}\nonumber
\\
\fl\qquad\,=\sum_{\vec{k}}\left(\frac{\hbar^2\vec{k}^2}{2 m_{e}}f^{e
} _ { \vec { k } }
+\frac{\hbar^2\vec{k}^2}{2
m_{h}}f^{h}_{\vec{k}}\right)-\frac{1}{2}\sum_{\vec{k},\vec{k}'}V_{\vec{k} -\vec { k }
'
}
\left(f^{e}_{\vec{k}}f^{e}_{\vec{k}'}+f^{h}_{\vec{k}}f^{h}_{\vec{k}'}
\right)-\sum_{\vec{k},\vec{k}'}V_{\vec{k}-\vec{k}'}P^{\star}_{\vec{k}}P_{\vec{k}'}
 \nonumber \\
\fl\qquad\quad+\frac{1}{2}\sum_{\vec{k},\vec{k}',\vec{q}}\left[V_{\vec{q}}\left(c^{\vec{q},\vec{k}',
\vec{k}}_{v ,
v;v,v}+c^{\vec{q},\vec{k}',\vec{k}}_{
c,c;c,c}\right)-2\,V_{\vec{k}'+\vec{q}-\vec{k}}\, c^{\vec{q},\vec{k}',\vec{k}}_{eh}\right
]\,,
\end{eqnarray}
which is an exact result for homogeneous excitation conditions. Using the cluster expansion
\cite{Kira:06b}, we identified the
incoherent two-particle correlations
\begin{eqnarray}
\label{eq:CE_corr}
 &&c_{v,v;v,v}^{\vec{q},\vec{k}',\vec{k}}\equiv\Delta\ev{a^{\dagger}_{v,\vec{k}}a^{\dagger}_{v,\vec{
k
}'}a_{v,\vec{k}'+\vec{q}}a_{v,\vec{k}-\vec{q}}}\,, \quad
 c_{c,c;c,c}^{\vec{q},\vec{k}',\vec{k}}\equiv\Delta\ev{a^{\dagger}_{c,\vec{k}}a^{\dagger}_{c,\vec{k
}'}a_{c,\vec{k}'+\vec{q}}a_{c,\vec{k}-\vec{q}}}\,, \nonumber \\
&&c_{eh}^{\vec{q},\vec{k}',\vec{k}}\equiv\Delta\ev{a^{\dagger}_{c,\vec{k}}a^{\dagger}_{v,\vec{
k
} ' } a_ {
c,\vec{k}'+\vec{q}}a_{v,\vec{k}-\vec{q}}}\,, 
\end{eqnarray}
which represent the truly correlated parts of the respective two-particle expectation
value. The first two correlations correspond to hole--hole and electron--electron correlations,
respectively. Electron--hole correlations are described by $c_{eh}^{\vec{q},\vec{k}',\vec{k}}$
where $\hbar\,\vec{q}$ defines the center-of-mass momentum of the correlated electron--hole pairs.

The only coherent quantity in equation~(\ref{eq:Emb}) is the microscopic
polarization
\begin{equation}
 \label{eq:pol}
P_{\vec{k}}\equiv
\ev{a^{\dagger}_{v,\vec{k}}a_{c,\vec{k}}}\,.
\end{equation}
Consequently, the average carrier energy $E_{\mrm{MB}}$ of any $\hat{\rho}_{\mrm{MB}}$ is
determined entirely by the single-particle expectation values $f^{\lambda}_{\vec{k}}$ and
$P_{\vec{k}}$ and the incoherent two-particle correlations $c^{\vec{q},\vec{k}',\vec{k}}$. In other
words, the system energy is \textit{directly} influenced by contributions up to second-order
correlations. We will show in section~\ref{Sec:pair_exc} that this fundamental property allows us to
determine the pair-excitation energetics of a given state when we know its singlets and doublets.
In other words, we do not need to identify the properties of the higher order clusters to compute the
pair-excitation energetics.

Since we are interested in long-living quasi-particles in the incoherent regime, we consider only those states $\hat{\rho}_{\mrm{MB}}$ which have vanishing coherences\cite{Book:11}. Therefore, we set $P_{\vec{k}}$ and all coherent correlations to zero from now on.
Furthermore, we assume conditions where the electron--hole correlations
$c^{\vec{q},\vec{k}',\vec{k}}_{eh}$ have a vanishing center-of-mass momentum
$\hbar\,\vec{q}=0$, i.e. we assume that the correlated pairs are at rest. As a result, the
electron--hole correlations can be expressed in terms of
\begin{equation}
 \label{eq:eq_com}
c^{\vec{q},\vec{k}',\vec{k}}_{eh}=\delta_{\vec{q},0}\,c^{\vec{q},\vec{k}',\vec{k}}_{eh}
\equiv\delta_{\vec{q},0}\,g_{\vec{k},\vec{k}'}\,.
\end{equation}
For homogeneous and incoherent
excitation conditions, the pair-correlation function can be written as
\begin{equation}
 \label{eq:pair_corr_fct}
g(\vec{r})\equiv\langle\hat{\Psi}^{\dagger}_{e}(\vec{r})\hat{\Psi}^{\dagger}_{h}(0)\hat{
\Psi } _ {h}
(0)\hat{\Psi}_{e}(\vec{r})\rangle=\rho_{e}\rho_{h}+\Delta g(\vec{r})\,, 
\end{equation}
compare equation~(\ref{eq:field}) \cite{Kira:06b}. The term 
$\rho_{e}\rho_{h}$ describes an uncorrelated electron--hole plasma contribution, whereas
the quasi-particle clusters determine the correlated part
\begin{equation}
 \label{eq:pc_corr}
\Delta g(\vec{r})=\frac{1}{S^2}\sum_{\vec{k},\vec{k}',\vec{q}}
c^{\vec{q},\vec{k}',\vec{k}}_{eh}\,\e^{\mrm{i}(\vec{k}'+\vec{q}-\vec{k})\cdot\vec{r}}=\frac{1}
{S^2}\sum_{\vec{k},\vec{k}'}g_{\vec{k},\vec{k}'}\,\e^{\mrm{i}(\vec{k}'-\vec{k})\cdot\vec{r}}\,.
\end{equation}
To describe e.g. excitons and similar
quasi-particles, we use an ansatz
\begin{equation}
 \label{eq:delta_g}
\Delta g(\vec{r})=|g_0\,\phi(\vec{r})|^2\,,
\end{equation}
where $g_0$ defines the strength of the correlation while the specific properties of the quasi-particles determine the normalized wavefunction $\phi(\vec{r})$. In order to compute the quasi-particle
energetics, we need to express $\Delta g(\vec{r})$ in terms of the electron--hole correlation
$g_{\vec{k},\vec{k}'}$. By writing
$\phi(\vec{r})=\frac{1}{S}\sum_{\vec{k}}\phi_{\vec{k}}\,\e^{\mrm{i}\vec{k}\cdot\vec{r}}$, we find the unique connection
\begin{equation}
\label{eq:gkk}
g_{\vec{k},\vec{k}'}=g_0^2\,\phi^{\star}_{\vec{k}}\,\phi_{\vec{k}'}\,,
\end{equation}
where $\phi(\vec{k})$ is the Fourier transformation of the wave function $\phi(\vec{r})$.

As shown in \ref{sec:cons}, the electron and hole distributions $f_{\vec{k}}^{e}$ and
$f_{\vec{k}}^{h}$, together with the incoherent correlations
$g_{\vec{k},\vec{k}'}$, $c^{\vec{q},\vec{k}',\vec{k}}_{v,v;v,v}$, and
$c^{\vec{q},\vec{k}',\vec{k}}_{c,c;c,c}$ must satisfy the general conservation laws
\begin{equation}
\label{eq:cons2} \fl
 \left(f^{e}_{\vec{k}}-\textstyle{\frac{1}{2}}\right)^2+g_{\vec{k},\vec{k}}-\sum\limits_{\vec
{ k}'}c^{ 0 , \vec { k } ' , \vec { k } } _{c,c;c,c}= \textstyle{\frac { 1
} { 4 }}  \,, \qquad 
 \left(f^{h}_{\vec{k}}-\textstyle{\frac{1}{2}}\right)^2+g_{\vec{k},\vec{k}}-\sum\limits_{\vec{
k } ' } c^{ 0 , \vec { k } ' , \vec { k } } _ {v,v;v,v}= \frac { 1 
}{ 4 }  \,.
\end{equation}
As a consequence, we have to connect $f^{e}_{\vec{k}}$ and $f^{h}_{\vec{k}}$ with $g_{\vec{k},\vec{k}}$,
$c^{\vec{q},\vec{k}',\vec{k}}_{c,c;c,c}$, and $c^{\vec{q},\vec{k}',\vec{k}}_{v,v;v,v}$
to have a self-consistent description of the many-body state. Therefore, equation~(\ref{eq:cons2})
has a
central role when the energetics of many-body states is solved self-consistently. 

We show in section~\ref{Sec:ee_corr} that the effect of electron--electron and hole--hole
correlations
can be neglected when the energetics of new quasi-particle states is analyzed. Therefore, we set
$c^{\vec{q},\vec{k}',\vec{k}}_{c,c;c,c}$ and $c^{\vec{q},\vec{k}',\vec{k}}_{v,v;v,v}$ to
zero such that equation~(\ref{eq:cons2}) reduces to
\begin{equation}
\label{eq:cons_final}
\left(f_{\vec{k}}-\textstyle{\frac{1}{2}}\right)^2+g_{\vec{k},\vec{k}}=\textstyle{\frac{1}{4}}\,,
\qquad f_{\vec{k}}\equiv
f_{\vec{k}}^{e}=f^{h}_{\vec{k}}\,.
\end{equation}
From this result, we see that the electron and hole distributions become identical as long
as correlations are dominated by $g_{\vec{k},\vec{k}'}$. A more general case with carrier--carrier
correlations is studied in section~\ref{Sec:ee_corr}. In the actual quasi-particle calculations,
we solve
equation~(\ref{eq:cons_final})
\begin{equation}
\label{eq:fk_gkk}
f_{\vec{k}}=\textstyle{\frac{1}{2}}\left(1\pm\sqrt{1-4\,g_{\vec{k},\vec{k}}}\right)\,,
\end{equation}
that limits $g_{\vec{k},\vec{k}}$ to be below $\textstyle{\frac{1}{4}}$. In other words, the
maximum of $g_0|\phi(\vec{k})|$ is $\textstyle{\frac{1}{2}}$, based on the connection (\ref{eq:gkk}).
The $``+``$ branch in equation~(\ref{eq:fk_gkk}) describes an inverted many-body system
$\hat{\rho}_{\mrm{MB}}$ corresponding to large electron--hole densities. Below inversion, only the
$``-``$
branch contributes.

Once the self-consistent pair $(f_{\vec{k}},g_{\vec{k},\vec{k}'})$ is found, we determine the
corresponding electron--hole density via
\begin{equation}
\label{eq:eh_dens}
\rho_{eh}=\frac{1}{S}\sum_{\vec{k}}f_{\vec{k}}\,,
\end{equation}
that becomes a functional of the electron--hole pair-correlation function due to its
$g_{\vec{k},\vec{k}'}$ dependence via equation~(\ref{eq:fk_gkk}). In sections~\ref{Sec:calc} and
\ref{Sec:Epair_drop}, we will
use equation~(\ref{eq:cons_final}) to self-consistently determine $f_{\vec{k}}$ and
$g_{\vec{k},\vec{k}'}$ for different quasi-particle configurations.

\subsection{\label{Sec:pair_exc} Pair-excitation energetics}

To generalize the Wannier equation (\ref{eq:Wannier}), we next analyze
the pair-excitation energetics of an arbitrary homogeneous initial state $\hat{\rho}_{\mrm{MB}}$. As
shown in section~\ref{sec:method}, the simplest class of pair excitations can be generated by using
the
coherent displacement-operator functional (\ref{eq:disp}).
The pair-excitation state is then given by
\begin{equation}
\label{eq:rho_g}
  \hat{\rho}[\psi]=\hat{D}[\psi]\,\hat{\rho}_{\mathrm{MB}}\, \hat{D}^{\dagger}[\psi]\,,
\end{equation}
which is properly normalized $\mrm{Tr}[\hat{\rho}[\psi]]=\mrm{Tr}[\hat{\rho}_{\mrm{MB}}]=1$,
as any density matrix should be.

As shown in \ref{app1}, the pair excitation generates the polarization and electron--hole
distribution
\begin{equation}
\label{eq:probe_p&f2} \fl
  P_{\vec{k},\psi}=
\left(1-f^{e}_{\vec{k}}-f^{h}_{\vec{k}}\right)\varepsilon\,\psi_{\vec{k}}+\mathcal{O}
(\varepsilon^3)\,, \quad
f_{\vec{k},\psi}=\left(1-f^{e}_{\vec{k}}-f^{h}_{\vec{k}}\right)\varepsilon^2\,
|\psi_{\vec{k}}|^2+\mathcal{O}(\varepsilon^3)\,,
\end{equation}
respectively, where we have applied the weak excitation limit $\varepsilon\ll 1$. For the sake of
completeness, we keep the explicit dependencies $f_{\vec{k}}^{e}$, $f^{h}_{\vec{k}}$, and
$c^{\vec{q},\vec{k}',\vec{k}}_{\lambda,\lambda;\lambda,\lambda}$ and take the limit of
dominant electron--hole correlation after the central results for the pair excitations have been
derived. In analogy to equation~(\ref{eq:pair_exc}), pair excitations add the average carrier
energy $E_{\mrm{pro}}[\psi]\equiv E[\psi]-E_{\mrm{MB}}$ to the system. Technically, $E[\psi]$ is
obtained by replacing $\rho_{\mrm{MB}}$ in equation~(\ref{eq:Emb}) by $\rho[\psi]$. The actual
derivation
is performed in \ref{app1}, yielding again an exact relation for incoherent quasi-particles:
\begin{eqnarray}
 \label{eq:Epro_ex} \fl
 E_{\mrm{pro}}[\psi]=\epsilon^2\sum_{\vec{k}}E_{\vec{k}}|\psi_{\vec{k}}|^2-\epsilon^2\sum_{\vec{k},
\vec{k}'}V^{\mrm{eff}}_{\vec{k},\vec{k}'}\,\psi_{
\vec { k } }\,\psi^{\star}_ { \vec { k }
'}
\nonumber \\ \fl \qquad\quad\,\,
 +\,\epsilon^2\sum_{\vec{k},\vec{k}',\vec{q}}V_{\vec{q}}\left(c^{\vec{q},\vec{k}',\vec{k}}_{v,v;v,v}
\,\psi_{\vec{k}-\vec{q}}\,\psi^ {
\star } _
{ \vec{k}}+c^{\vec{q},\vec{k}',\vec{k}}_{c,c;c,
c}\,
\psi_{\vec{k}}\,\psi^{\star}_{\vec{k}-\vec{q}}-\re
[c^{\vec{q},\vec{k}',\vec{k}}_{v,v;v,v}+c^{\vec{q},\vec{k}',\vec{k}}_{c,c;c,c}]|\psi_{
\vec { k } } |^2
\right) \nonumber \\ \fl \qquad\quad\,\,
 +\,\mathcal{O}(\varepsilon^3)\,,
\end{eqnarray}
where we identified the renormalized kinetic electron--hole pair energy
\begin{equation}
 \label{eq:E_renorm} \fl
 E_{\vec{k}}\equiv\left[\frac{\hbar^2\vec{k}^2}{2\mu}-\sum_{\vec{k}'}V_{\vec{k}
-\vec{k}'}\left(f^{e}_{\vec{k}'}+f^{h}_{\vec{k}'}\right)\right]\left(1-f^{e}_{\vec
{ k } }-f^{h}_{\vec
{ k } }
\right)+2\sum_{\vec{k}'}V_{\vec{k}-\vec{k}'}\,g_{\vec{k},\vec{k}'}\,.
\end{equation}
The unscreened Coulomb interaction $V_{\vec{k}-\vec{k}'}$ is modified through the presence of electron--hole densities and correlations via
\begin{equation}
\label{eq:VS}
V^{\mrm{eff}}_{\vec{k},\vec{k}'}\equiv
\left(1-f^{e}_{\vec{k}}-f^{h}_{\vec{k}}\right)
V_{\vec{k}-\vec{k}'}\left(1-f^{e}_{
\vec{k}'}-f^{h}_{
\vec{k}'}\right)+2g_{\vec{k},\vec{k}'}V_{\vec{k} -\vec{k}'}\,.
\end{equation}
Since the phase-space filling factor $(1-f^{e}_{\vec{k}}-f^{h}_{\vec{k}})$ becomes negative once
inversion is reached, the excitation level changes the nature of the effective electron--hole Coulomb interaction from attractive to
repulsive. At the same time, $g_{\vec{k},\vec{k}'}$ can either enhance
or decrease the Coulomb interaction depending on the nature of the pair correlation. The exact
generalization of equation~(\ref{eq:Epro_ex}) for coherent quasi-particles is presented in
\ref{Sec:Wannier_com}.

\subsection{Generalized Wannier equation}

As in section~\ref{sec:ord_Wannier}, we minimize the functional $E_{\mrm{pro}}[\psi]$ with the
constraint that the excitation $\varepsilon^2\sum_{\vec{k}}|\psi_{\vec{k}}|^2$ remains constant.
Following the same variational steps as those producing equation~(\ref{eq:Wannier}), we obtain the
\textit{generalized Wannier equation} for incoherent quasi-particles:
\begin{eqnarray}
 \label{eq:gen_Wannier} 
 E_{\vec{k}}\psi_{\vec{k}}-\sum_{\vec{k}'}V^{\mrm{eff}}_{\vec{k},\vec{k}'}\psi_{\vec{k}}+\sum_{\vec{
k}',\vec{q}}V_{\vec{q}}\left(c^{\vec{q},\vec{k}',\vec{k}+\vec{q}}_{c,c;c,c}\,\psi_{\vec{
k }
+\vec{q}}+c^{\vec{q},\vec{k}',\vec{k}}_{v,v;v,v}\,\psi_{\vec{k}-\vec{q}}\right) \nonumber \\
+\sum_{\vec{k}',\vec{
q}}V_{\vec{q}}\,\re\left[c^{\vec{q},\vec{k}',\vec{k}}_{c,c;c,c}+c^{\vec{q},\vec{k}',\vec{k}}_{
v,v;v,v}\right]\psi_{\vec{k}}=E_{\lambda}\psi_{\vec{k}}\,.
\end{eqnarray}
For vanishing electron--hole densities and correlations, equation~(\ref{eq:gen_Wannier}) reduces to
the
ordinary exciton Wannier equation (\ref{eq:Wannier}). Since the presence of two-particle correlations
and densities modifies the effective Coulomb interaction, it is possible that new quasi-particles emerge. The generalized Wannier equation with all coherent and incoherent contributions is
presented in \ref{Sec:Wannier_com}.

For the identification of the quasi-particle energy, we compute the energy per excited
electron--hole pair (\ref{eq:E_per}). The number of excited electron--hole pairs of the probed
many-body system is
\begin{equation}
 \label{eq:Npro2}
N_{\mrm{pro}}\equiv\sum_{\vec{k}}f_{\vec{k},\psi}=\varepsilon^2\sum_{\vec{k}}\left(1-f^{e}_{
\vec
{ k } } -f^ {h}_{\vec{k}}\right)|\psi_{\vec{k}}|^2\,, 
\end{equation}
according to equation~(\ref{eq:probe_p&f2}). By inserting equation (\ref{eq:gen_Wannier}) into
equation~(\ref{eq:Epro_ex}) and using the definitions (\ref{eq:E_per}) and (\ref{eq:Npro2}), the
energy per excited electron--hole pairs follows from
\begin{equation}
 \label{eq:Epp_lambda}
 \bar{E}_{\mrm{pro}}=E_{\lambda}\frac{\sum_{\vec{k}}|\psi_{\vec{k}}|^2}{\sum_{\vec{k}}|\psi_{\vec{
k } }|^2\left(1-f^{e}_{\vec{k}}-f^{h}_{\vec{k}}\right)}\,,
\end{equation}
that defines the quasi-particle energy, based on the discussion in section~\ref{sec:method}

\section{\label{Sec:calc} Pair-excitation spectrum of the degenerate Fermi gas and of incoherent excitons}

For all our numerical evaluations, we use the parameters of a typical $10$\,nm GaAs-QW system. Here, the reduced mass is $\mu=0.0581\,m_0$ where $m_0$ is the free-electron mass and the
$1s$-exciton binding energy is $E_{\mrm{B}}=9.5$\,meV. This is obtained by using the dielectric
constant $\varepsilon_{\mrm{r}}=13.74$ of GaAs in the Coulomb interaction.

To compute the quasi-particle energetics for a given electron--hole density $\rho_{eh}$, we
always start from the conservation law (\ref{eq:cons_final}) to generate a self-consistent many-body
state $\hat{\rho}_{\mrm{MB}}$. We then use the found self-consistent
pair $(f_{\vec{k}},g_{\vec{k},\vec{k}'})$ as an input to the generalized Wannier equation
(\ref{eq:gen_Wannier}) and numerically solve the pair excitation $\psi_{\vec{k}}$ and $E_{\lambda}$.
As shown in section~\ref{Sec:ee_corr}, the effect of electron-electron and hole--hole
correlations
on the quasi-particle energetics is negligible such that we set
$c^{\vec{q},\vec{k}',\vec{k}}_{c,c;c,c}$ and $c^{\vec{q},\vec{k}',\vec{k}}_{v,v;v,v}$ to zero in
equation~(\ref{eq:gen_Wannier}).

The variational computations rigorously determine only the lowest energy $E_0$.
However, it is useful to analyze also the characteristics of the excited states $E_{\lambda}$ to
gain additional information about the energetics of the pair excitation acting upon
$\hat{\rho}_{\mrm{MB}}$. To deduce the quasi-particle energetics, we normalize the energy
$E_{\lambda}$ via equation~(\ref{eq:Epp_lambda}). The resulting energy per excited electron--hole
pair
$\bar{E}_{\mrm{pro}}$ defines then the detectable energy resonances.

\subsection{Degenerate Fermi gas}

The simplest form of $\hat{\rho}_{\mrm{MB}}$ for an excited state is provided by the degenerate
Fermi gas\cite{DeMarco:99,Holland:01,O'Hara:02,Greiner:03}
\begin{equation}
 \label{eq:Fermi}
 f_{\vec{k}}=\theta(k-k_{\mrm{F}})\,,\qquad g_{\vec{k},\vec{k}'}=0\,,
\end{equation}
because the two-particle correlations vanish. It is straight forward to show that the pair
$(f_{\vec{k}},g_{\vec{k},\vec{k}'})$ satisfies the conservation law (\ref{eq:cons_final}) even
though the system is inverted for all $k$ below the Fermi wave vector
$k_{\mrm{F}}=\sqrt{4\pi\rho_{eh}}$. Due to this inversion, the degenerate Fermi gas provides a
simple model to study quasi-particle excitations under optical gain conditions.

\begin{figure}[t]
\includegraphics*[scale=0.45]{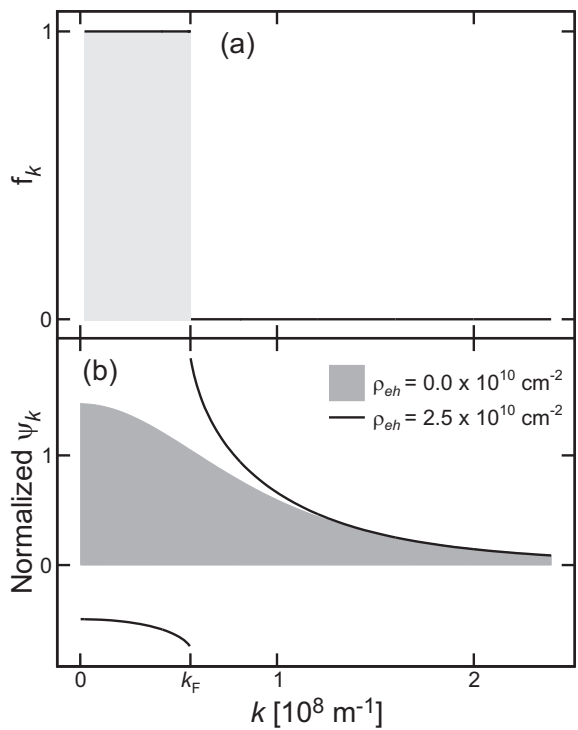}
\centering
\caption{Solutions of the generalized Wannier equation for degenerate Fermi gas.
(a) The electron--hole distribution $f_{\vec{k}}$ is shown as function of $k$ for 
$\rho_{eh}=2.5\times 10^{10}\,\mrm{cm}^{-2}$ and $k_{\mrm{F}}=0.56\times
10^8\,\mrm{m}^{-1}$. (b) Normalized ground-state wavefunction
$\psi_{\vec{k}}$ for vanishing electron--hole density (shaded area) and $\rho_{eh}=2.5\times
10^{10}\,\mrm{cm}^{-2}$ (solid line).}
\label{Fig2}
\end{figure}

Figure \ref{Fig2}(a) presents the
electron--hole distribution $f_{\vec{k}}$ as function of $k$ for the electron--hole density
$\rho_{eh}=2.5\times 10^{10}\,\mrm{cm}^{-2}$. The distribution has a Fermi edge at
$k_{\mrm{F}}=0.56\times 10^8\,\mrm{m}^{-1}$ while $g_{\vec{k},\vec{k}}$ is zero for all $k$ values
(not shown). The numerically computed ground-state wave function
$\psi_{\vec{k}}$ is plotted in figure~\ref{Fig2}(b) as solid line. We have applied the
normalization $\sum_{\vec{k}}|\psi_{\vec{k}}|^2=1$.  As a comparison, we also show the
corresponding zero-density result $(f_{\vec{k}}=0,g_{\vec{k},\vec{k}'}=0)$ as shaded area. While the
zero-density wave function decays monotonically from the value 1.47, the
degenerate Fermi gas has a $\psi_{\vec{k}}$ that is negative-valued up to the Fermi
edge $k_{\mrm{F}}$. Exactly at $k=k_{\mrm{F}}$, $\psi_{\vec{k}}$ abruptly jumps from the value
-0.74 to 1.89. Above roughly $k=1.3\times10^{8}\,\mrm{m}^{-1}$, both wave functions
show a similar decay. The energetics of the related pair excitations is discussed later in
section~\ref{sec:energetics1}.

\subsection{Incoherent excitons}
\label{Sec:X}

\begin{figure}[t]
\includegraphics*[scale=0.45]{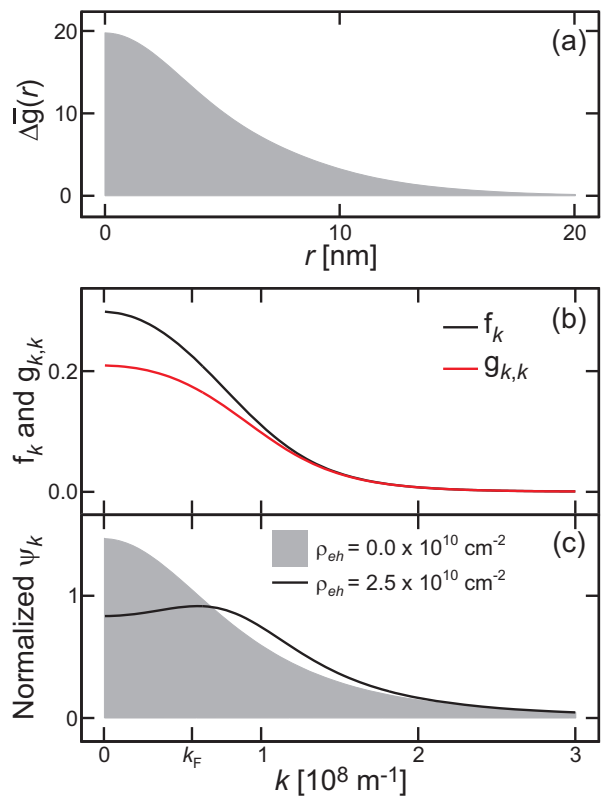}
\centering
\caption{Solutions of the generalized Wannier equation for incoherent excitons.
(a) The normalized electron--hole pair-correlation function $\Delta\bar{g}(r)$ is
shown for $\rho_{eh}=2.5\times 10^{10}\,\mrm{cm}^{-2}$. (b) The
corresponding electron--hole distribution $f_{\vec{k}}$ (black line) and the correlation
$g_{\vec{k},\vec{k}}$ (red line) as function of $k$. (c) Normalized ground-state
wavefunction for vanishing electron--hole density (shaded area) and $\rho_{eh}=2.5\times
10^{10}\,\mrm{cm}^{-2}$ (solid line)}
\label{Fig3}
\end{figure}

According to the ansatz (\ref{eq:gkk}), the exciton state is determined by the electron--hole
pair-correlation function
\begin{equation}
  \label{eg:gkk_1s}
  g_{\vec{k},\vec{k}'}=\phi_{1s,\vec{k}}\phi_{1s,\vec{k}'}\,,
\end{equation}
with the $1s$-exciton wavefunction $\phi_{1s,\vec{k}}$ defining the initial many-body state
$\hat{\rho}_{\mrm{MB}}$, not the pair-excitation state. Here, we have included the strength of
the electron--hole correlation $g_0$ into the $1s$-exciton wavefunction to simplify the notation. To
compute $\phi_{1s,\vec{k}}$, we have to solve the ordinary density-dependent Wannier equation
\cite{Book:11,Kira:06b}
\begin{equation}
  \label{eq:Wannier_1s} \fl
\tilde{E}_{\vec{k}}\,\phi_{1s,\vec{k}}-\left(1-2\,f_{\vec{k}}\right)\sum_{\vec{k}'}V_{\vec{k}-\vec{
k}'}\,\phi_{1s,\vec{k}'}=E_{1s}\,\phi_{1s,\vec{k}}\,, \quad
\tilde{E}_{\vec{k}}=\frac{\hbar^2\vec{k}^2}{2\mu}-2\sum_{\vec{k}'}V_{\vec{k}-\vec{k}'}f_{\vec{k}'}\,
,
\end{equation}
with the constraint imposed by the conservation law (\ref{eq:cons_final}). In practice, we solve
equations~(\ref{eq:cons_final}) and (\ref{eq:Wannier_1s}) iteratively. 
Since the specific choice $E_{1s}$ defines the electron--hole density
(\ref{eq:eh_dens}) uniquely, we can directly identify the self-consistent pair
$(f_{\vec{k}},g_{\vec{k},\vec{k}'})$ as function of
$\rho_{eh}$. The explicit steps of the iteration cycle are presented in \ref{Sec:X_solver}.

Figure \ref{Fig3}(a) shows the resulting normalized electron--hole pair-correlation
function $\Delta\bar{g}(r)\equiv\Delta g(r)/\rho_{eh}^2$ 
for an electron--hole density of $\rho_{eh}=2.5\times 10^{10}\,\mrm{cm}^{-2}$. For the incoherent excitons, 
$\Delta\bar{g}(r)$ is a monotonically decaying function. The
corresponding iteratively solved $f_{\vec{k}}$ (black line) and $g_{\vec{k},\vec{k}}$ (red line) are
plotted in figure~\ref{Fig3}(b). The pair correlation $g_{\vec{k},\vec{k}}$ decays
monotonically from the value 0.21. Also the electron--hole distribution
$f_{\vec{k}}$ function decreases monotonically, peaking at 0.30. This implies that the
phase-space filling already reduces the strength of the effective Coulomb potential
(\ref{eq:VS}) for small momentum states which typically dominate the majority of ground-state
configurations. 

The corresponding normalized ground-state wavefunction $\psi_{\vec{k}}$ of the pair excitation is
shown in figure~\ref{Fig3}(c) (solid line) together with the zero-density result (shaded area).
Both
functions show a similar decay for $k$ values larger than $2\times
10^8\,\mrm{m}^{-1}$. In contrast to the zero-density result, we observe that $\psi_{\vec{k}}$ has a
peak at $k=0.59\times 10^8\,\mrm{m}^{-1}$. Interestingly, the maximum of $\psi_{\vec{k}}$ is
close to $k_{\mrm{F}}$ of the degenerate Fermi gas analyzed in figure~\ref{Fig2} because both
cases have the same density giving rise to sufficiently strong phase-space filling effects.

\begin{figure}[t]
\includegraphics*[scale=0.5,angle=-90]{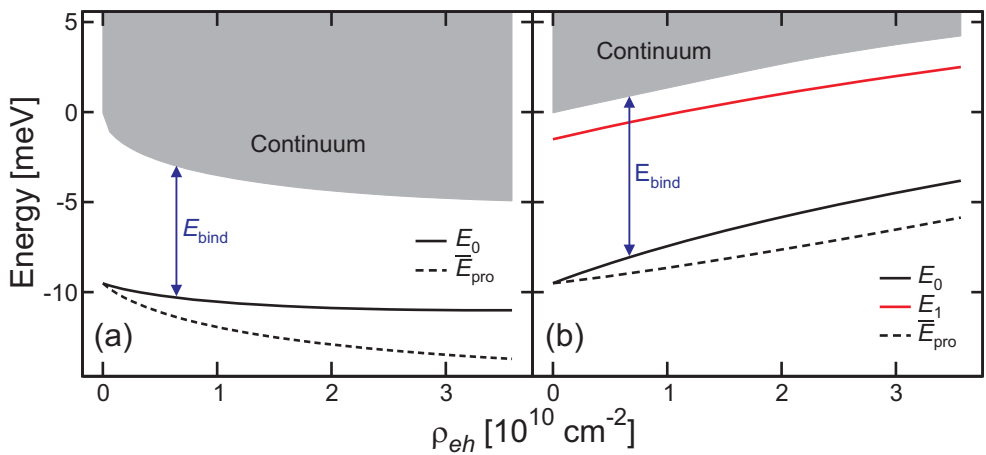}
\centering
\caption{Pair-excitation energetics for the degenerate Fermi gas vs. incoherent excitons.
(a) The ground-state energy $E_0$ (black solid line), the continuum (shaded are), and the
energy per excited electron--hole pair $\bar{E}_{\mrm{pro}}$ (dashed line) are presented as function
of the electron--hole density $\rho_{eh}$ for the degenerate Fermi gas. The same analysis is plotted in
(b) for the exciton state. Additionally, the red solid line shows the energy of the
first excited state $E_1$.}
\label{Fig4}
\end{figure}

\subsection{\label{sec:energetics1} Energetics of pair excitations}

We next analyze the influence of the electron--hole density $\rho_{eh}$ on the pair-excitation
energetics for the degenerate Fermi gas and for incoherent excitons. The result for the degenerate Fermi gas
is presented in figure~\ref{Fig4}(a) where the ground-state energy $E_{0}$ (solid line), the
continuum (shaded area), and the ground-state energy per excited electron--hole pair
$\bar{E}_{\mrm{pro}}$ (dashed line) are plotted as function of $\rho_{eh}$. We see that the energy difference between
$E_0$ and the ionized states is considerably reduced from $9.5$\,meV to $6.1$\,meV as the density is
increased from zero to $\rho_{eh}=3.6\times 10^{10}\,\mrm{cm}^{-2}$. This decrease is already an
indication that non of the excited states remain bound for elevated densities. At the same
time, the ground-state energy shows only a slight red shift while the continuum is strongly red
shifted such that the first excited state becomes ionized for electron--hole densities above
$\rho_{eh}=2\times 10^{9}\,\mrm{cm}^{-2}$. The detectable pair-excitation energy is defined by
$\bar{E}_{\mrm{pro}}$, according to equation~(\ref{eq:Epp_lambda}). As a general trend,
$\bar{E}_{\mrm{pro}}$ is slightly smaller than $E_0$. We also observe that
$\bar{E}_{\mrm{pro}}$ remains relatively stable as the density is increased. This implies that the
semiconductor absorption and gain peaks appear at roughly the same position independent of
electron--hole density. This conclusion is consistent with fully microscopic absorption
\cite{Smith:10} and gain calculations \cite{Gerhardt:04,Koukourakis:12} and measurements
\cite{Ellmers:98,Hofmann:02}.

The pair-excitation energetics of the exciton state (\ref{eg:gkk_1s})--(\ref{eq:Wannier_1s})
is presented in figure~\ref{Fig4}(b) for the initial exciton state analyzed in
figure~\ref{Fig3}. The black line compares the ground state $E_0$ with the first excited
state $E_1$ (red line) while the shaded area indicates the ionized solutions. In contrast
to the degenerate Fermi gas, the ground-state energy blue shifts. This blue shift remains present in
$\bar{E}_{\mrm{pro}}$ (dashed line) and is consistent with the blue shift of the excitonic
absorption when excitons are present in the system, as detected in several measurements
\cite{Khitrova:99,Smith:10,Peyghambarian:84,Kira:11}. In particular, $E_0$ blue shifts
faster than the continuum does. If we interpret the energy difference of $E_0$ and continuum as the
exciton-binding energy, we find that the exciton-binding energy decreases from $9.5$\,meV to
$8.0$\,meV as the density is increased to $\rho_{eh}=3.6\times 10^{10}\,\mrm{cm}^{-2}$, which shows
that excitons remain bound even at elevated densities. For later reference, the density $2.5\times
10^{10}\,\mrm{cm}^{-2}$ produces $\bar{E}_{\mrm{pro}}=-7.1$\,meV energy per excited
electron--hole pair.

\section{Pair-excitation spectrum of quantum droplets}
\label{Sec:Epair_drop}

To define a quantum droplet state, we assume that the electron--hole pairs form a
liquid confined within a small droplet with a radius $R$ as discussed in connection with
figure~\ref{Fig1}. Since the QW is two dimensional, the
droplet is confined inside a circular disc with radius $R$. We assume that the droplet has a hard
shell created by the Fermi pressure of the plasma acting upon the droplet. As a result, the
solutions correspond to standing waves. Therefore, we define the quantum droplet state via the
standing-wave ansatz
\begin{equation}
  \label{eq:phi_drop}
  \phi(r)=J_0\left(x_n\textstyle{\frac{r}{R}}\right)\,\e^{-\kappa r}\theta(R-r)\,,
\end{equation}
to be used in equation~(\ref{eq:delta_g}). Here, $x_n$ is the $n$-th zero of the Bessel function
$J_0(x)$. The Heaviside $\theta(x)$ function confines the droplet inside a
circular disk with radius $R$. The additional decay constant $\kappa$ is used for adjusting the
electron--hole density (\ref{eq:eh_dens}) when the quantum droplet has radius $R$ and $n$ rings.

For a given quantum droplet radius $R$, ring number $n$, and electron--hole density $\rho_{eh}$, we
fix
the peak amplitude of $g_{\vec{k},\vec{k}}$ to $g_{\mrm{max}}=\max[g_{\vec{k},\vec{k}}]$ which
defines the strength of the electron--hole
correlations. This settles $g_0$ for any given $(R,n,\rho_{eh})$ combination. Based on the
discussion following equation~(\ref{eq:fk_gkk}), the largest possible peak amplitude of
$g_{\vec{k},\vec{k}}$ is $\textstyle{\frac{1}{4}}$ which yields vanishing $(1-2f_{\vec{k}})$ at the
corresponding momentum.

Once $g_0$ produces a fixed $g_{\mrm{max}}$, we only need to find which $\kappa$ value produces the
correct density for a given $(R,n)$ combination. In other words, $\kappa$ alters $\rho_{eh}$
because it changes the width of $g_0\,\phi_{\vec{k}}$ whose peak amplitude is already fixed. Since
we want to solve $\bar{E}_{\mrm{pro}}$ for a given
$(R,n,\rho_{eh})$ combination, we solve the specific $\kappa$ value iteratively. In more
detail, we construct $f_{\vec{k}}$ by using $g_0\,\phi_{\vec{k}}$ as input to
equation~(\ref{eq:fk_gkk})
for a fixed $(R,n)$ as function of $\kappa$. We then find iteratively which $\kappa$ satisfies the
density condition (\ref{eq:eh_dens}).

\begin{figure}[t]
\includegraphics*[scale=0.45]{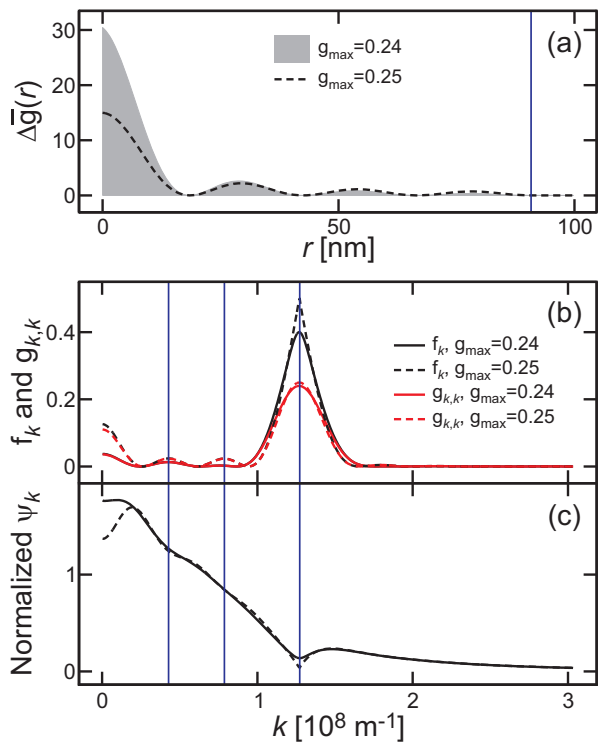}
\centering
\caption{Solutions of the generalized Wannier equation for quantum droplets.
(a) The normalized electron--hole pair-correlation function $\Delta
\bar{g}(r)$ is shown for $g_{\mrm{max}}=0.24$
(shaded area) and $g_{\mrm{max}}=\textstyle{\frac{1}{4}}$ (dashed line). The quantum
droplet has $n=4$ rings, $R=90.8$\,nm (vertical line), and $\rho_{eh}=2.5\times
10^{10}\,\mrm{cm}^{-2}$. (b)
The corresponding electron--hole distribution $f_{\vec{k}}$ (black lines) and correlation
$g_{\vec{k},\vec{k}}$ (red lines) as function of $k$ for $g_{\mrm{max}}=0.24$
(solid lines) and $g_{\mrm{max}}=\textstyle{\frac{1}{4}}$ (dashed lines). (c) The
resulting normalized ground-state wavefunctions $\psi_{\vec{k}}$.}
\label{Fig5}
\end{figure}

Figure \ref{Fig5}(a) presents the normalized electron--hole pair-correlation function $\Delta
\bar{g}(r)$ for an electron--hole correlation strength of $g_{\mrm{max}}=0.24$ (shaded
area) and $g_{\mrm{max}}=\textstyle{\frac{1}{4}}$ (dashed line). The quantum droplet has
$n=4$
rings and a radius of $R=90.8$\,nm indicated by a vertical line. We assume that the
electron--hole density is $\rho_{eh}=2.5\times 10^{10}\,\mrm{cm}^{-2}$ such that the iteration
yields $\kappa=2.2\times 10^7\,\mrm{m}^{-1}$ ($\kappa=3.4\times 10^6\,\mrm{m}^{-1}$) for
$g_{\mrm{max}}=0.24$ ($g_{\mrm{max}}=\textstyle{\frac{1}{4}}$), which
settles the consistent quantum droplet configuration. We observe that $\Delta \bar{g}(r)$ has four
rings
including the half oscillation close to the origin which appears due to the Coulomb attraction
between electrons and holes. Additionally, the electron--hole pair-correlation function is only
nonzero up to the hard shell at $r=R$, according to equation~(\ref{eq:phi_drop}). By comparing the
results of $g_{\mrm{max}}=0.24$ and $g_{\mrm{max}}=\textstyle{\frac{1}{4}}$,
we note that the oscillation amplitude decreases slower as function of $r$ with increasing
$g_{\mrm{max}}$ because the decay parameter $\kappa$ is smaller for elevated
$g_{\mrm{max}}$.

The corresponding self-consistently computed electron--hole distribution
$f_{\vec{k}}$ and correlation $g_{\vec{k},\vec{k}}$ are plotted in figure~\ref{Fig5}(b) as
black and red lines, respectively, for $g_{\mrm{max}}=0.24$
(solid lines) and $g_{\mrm{max}}=\textstyle{\frac{1}{4}}$ (dashed lines). The
electron--hole distribution $f_{\vec{k}}$ peaks to $0.4$
($0.5$) at $k=1.3\times 10^8\,\mrm{m}^{-1}$ for $g_{\mrm{max}}=0.24$
($g_{\mrm{max}}=\textstyle{\frac{1}{4}}$). We see that the peak of $f_{\vec{k}}$
sharpens as $g_{\mrm{max}}$ is increased. Interestingly, $f_{\vec{k}}$ and
$g_{\vec{k},\vec{k}}$ show small oscillations indicated by vertical lines whose amplitude becomes
larger with increasing electron--hole correlation strength. 

As we compare the $f_{\vec{k}}$ of the
quantum droplets with that of the excitons (figure~\ref{Fig3}(b)), we note that quantum
droplets exhibit a significant
reduction of the Pauli blocking, i.e. $(1-2f_{\vec{k}})$, at small momenta. As a result, quantum
droplets
produce a stronger electron--hole attraction than excitons for low $\vec{k}$, which makes the
formation of these quasi-particle states possible once the carrier density becomes large enough.
Figure \ref{Fig5}(c) presents the corresponding normalized ground-state wavefunctions
$\psi_{\vec{k}}$. The wavefunction $\psi_{\vec{k}}$ is qualitatively different from the state
obtained for both, the degenerate Fermi gas and excitons, presented in figures~\ref{Fig2}(b)
and \ref{Fig3}(c), respectively. In particular, the quantum droplet produces a
$\psi_{\vec{k}}$ that
has small oscillations for small $k$ (vertical lines) which are synchronized with the oscillations
of $f_{\vec{k}}$. Additionally, $f_{\vec{k}}$ shows a strong dip close to the inversion $k=1.3\times
10^8\,\mrm{m}^{-1}$. The dip becomes more pronounced as $g_{\mrm{max}}$ is increased. 

As discussed above, the largest possible peak amplitude of $g_{\vec{k},\vec{k}}$ is
$\textstyle{\frac{1}{4}}$. By approaching $g_{\mrm{max}}=\textstyle{\frac{1}{4}}$, the
energy per excited electron--hole pair $\bar{E}_{\mrm{pro}}$ decreases slightly from
$\bar{E}_{\mrm{pro}}=-10.12$\,meV to $\bar{E}_{\mrm{pro}}=-10.14$\,meV
as $g_{\mrm{max}}$ is changed from $0.24$ to $\textstyle{\frac{1}{4}}$. In
general, for a fixed quantum-droplet radius $R$, ring number $n$, and electron--hole density
$\rho_{eh}$, we find that $\bar{E}_{\mrm{pro}}$ is minimized when the amplitude of
$g_{\vec{k},\vec{k}}$ is maximized. Consequently, we use
$g_{\mrm{max}}=\textstyle{\frac{1}{4}}$ in our
calculations to study the energetics of quantum droplets. For this particular case, the quantum
droplet's ground state is $3.0$\,meV below the exciton energy, based on the analysis in
section~\ref{Sec:X}. Therefore, the quantum droplets are quasi-particles where electron--hole pairs
are
stronger bound than in excitons, as concluded above.

\begin{figure}[t]
\includegraphics*[scale=0.5,angle=-90]{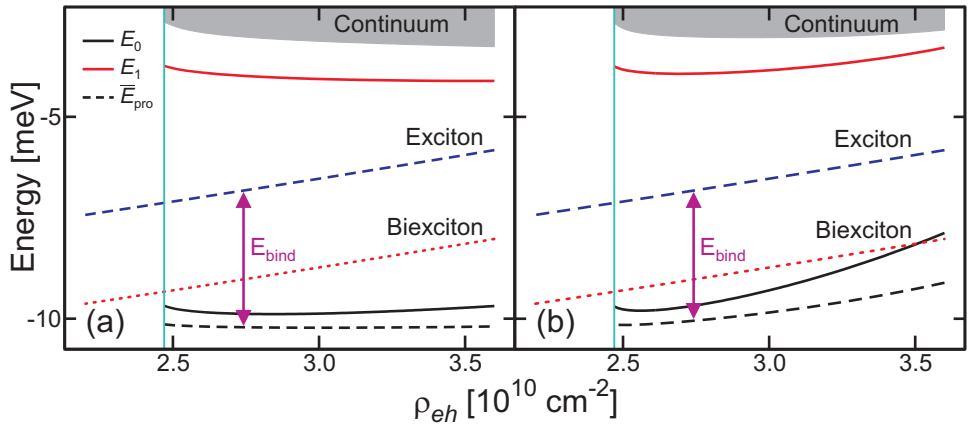}
\centering
\caption{Energetics of quantum droplets.
(a) The ground-state energy $E_0$ (black solid line), the first excited state $E_1$ (red solid
line), the continuum (shaded area), and the energy per excited electron--hole pair (black dashed
line) are presented as function of $\rho_{eh}$. The quantum droplet has $n=4$ rings
and $R=90.8$\,nm. The density-dependent exciton (dashed blue line) and biexciton-binding
energy (dotted red line) are also plotted. (b) The corresponding
result for quantum droplets with the density-dependent $R$ defined in equation~(\ref{eq:R_dens}).}
\label{Fig6}
\end{figure}

\subsection{Density dependence}
\label{Sec:drop_r}

The quantum droplet ansatz (\ref{eq:phi_drop}) is based on a postulated radius $R$ for the
correlation
bubble. Even though we find the self-consistent configuration $(f_{\vec{k}},g_{\vec{k},\vec{k}})$ 
for each $R$, we still need to determine the stable quantum droplet configurations.
As the main condition, the quantum droplet's pair-excitation energy must be lower than that of the
excitons and the biexcitons.

In the formation scheme of macroscopic electron--hole droplets, these droplets emerge only after a
critical density is exceeded
\cite{Jeffries:75}. In addition, stable droplets grow in size as the overall particle density is
increased. Therefore, it is reasonable to assume that also quantum droplets share these properties.
We use the simplest form where the area of the quantum droplet scales linearly with density. This
condition connects the radius and density via
\begin{equation}
\label{eq:R_dens}
 R=R_0\sqrt{\frac{\rho_{eh}}{\rho_0}}\,,
\end{equation}
where $R_0$ is the radius at reference density $\rho_0$. To determine the effect of the droplet's
$\rho_{eh}$-dependent size, we also compute the quantum droplet properties for a fixed $R=R_0$. In
the
actual calculations, we use $R_0=90.8$\,nm and $\rho_0=2.5\times 10^{10}\,\mrm{cm}^{-2}$. 

In both cases, we find the fully consistent pair $(f_{\vec{k}},g_{\vec{k},\vec{k}'})$ as described
in section~\ref{Sec:Epair_drop} and compute the pair-excitation energy for different $\rho_{eh}$.
Figure \ref{Fig6}(a) shows the ground-state energy $E_0$ (solid black line), the first
excited state $E_1$ (solid red line), the continuum (shaded area), and the energy per excited
electron--hole pair (black dashed line) as function of $\rho_{eh}$ when a constant-$R$ quantum
droplet has
$n=4$ rings. The corresponding result for the density-dependent $R$,
defined by equation~(\ref{eq:R_dens}), is shown in figure~\ref{Fig6}(b). In both frames, the
position of the density-dependent exciton (dashed blue line) and biexciton energy (dotted red line)
are indicated, based on the calculation shown in figure~\ref{Fig4} and the experimentally
deduced biexciton binding energy 2.2\,meV in reference~\cite{Kira:11}.

For both $R$ models, the quantum droplet's pair-excitation energy $\bar{E}_{\mrm{pro}}$ (black
dashed
line) is significantly lower than both the exciton and the biexciton energy, which makes the
$(n=4)$-ring
quantum droplet energetically stable for densities exceeding $\rho_{eh}=2.5\times
10^{10}\,\mrm{cm}^{-2}$.
We also see that all excited states of the quantum droplets have a higher energy than the exciton.
Therefore,
only the quantum droplet's ground state is energetically stable enough to exist permanently.
However, the
quantum droplet state with $n=4$ rings does not exist for an electron--hole density below
$\rho_{eh}=2.47\times 10^{10}\,\mrm{cm}^{-2}$ (vertical line) because this case corresponds
to the smallest possible $\kappa=0$. In other words, one cannot lower $\kappa$ to make
$f_{\vec{k}}$ narrower in order to produce $\rho_{eh}$ smaller than
$2.47\times10^{10}\,\mrm{cm}^{-2}$. More generally, one can compute the threshold $\rho_{eh}$ of a
quantum droplet with $n$ rings by setting $\kappa$ to zero in equation~(\ref{eq:phi_drop}) and by
generating
the corresponding $\phi_{\vec{k}}$, $g_{\vec{k},\vec{k}}$, and $f_{\vec{k}}$ via
equation~(\ref{eq:cons_final}). Since $\phi_{\vec{k}}$ and $f_{\vec{k}}$ peak at $k$ that is
proportional
to $x_n$, it is clear that $\rho_{eh} \propto \int_{0}^{\infty}\mrm{d}k\,kf_{\vec{k}}$ increases
monotonically as function of $n$. Therefore, one finds quantum droplets with a higher ring number
only at elevated densities.

\begin{figure}[t]
\includegraphics*[scale=0.5]{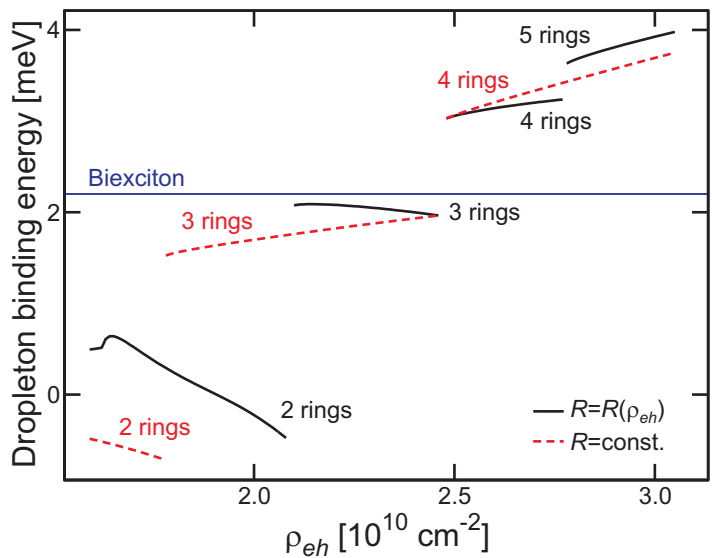}
\centering
\caption{Ground-state energy of quantum droplets. The ground-state energy is presented as function
of $\rho_{eh}$ for a constant (dashed line) and density-dependent $R$ (solid
line). The biexciton-binding energy is indicated by the horizontal line. }
\label{Fig7}
\end{figure}

\subsection{\label{Subsec:G_drop} Ground-state energy}

To determine the quantum droplet's binding energy, we define
\begin{equation}
  \label{eq:drop_bind}
  E_{\mrm{bind}}\equiv \bar{E}_{\mrm{pro}}(1s)-\bar{E}_{\mrm{pro}}(\mrm{droplet})\,,
\end{equation}
where $\bar{E}_{\mrm{pro}}(1s)$ and $\bar{E}_{\mrm{pro}}(\mrm{droplet})$ are
the ground-state energies of the exciton and the quantum droplet, respectively. Figure \ref{Fig7}
presents
$E_{\mrm{bind}}$ for all possible ring numbers for both constant $R$ (dashed line) and
$\rho_{eh}$-dependent $R$ (solid line), as function of $\rho_{eh}$. Here, we follow the
lowest $E_{\mrm{bind}}$ among all $n$-ring states as the ground state of the quantum droplet. As
explained in section~\ref{Sec:drop_r}, each $n$-ring state appears as an individual threshold
density is crossed. The horizontal line indicates the binding energy of the biexciton. We see
that both droplet-radius configurations produce discrete energy bands. As the electron--hole
density is increased, new energy levels appear as sharp transitions. Each transition increases the
ring number $n$ by one such that the ring number directly defines the quantum number for the
discrete energy levels. We see that only quantum droplets with more or equal than four rings have a
larger binding than biexcitons do, making 1-, 2-, and 3-ring quantum droplets instable.
The constant $R$ and the density-dependent $R$ produce a qualitatively similar energy structure. As
main differences, the constant $R$ produces ring-to-ring transitions at higher densities and the
energy bands spread to a wider energy range. For example, the energy range of the $n=4$ energy band
is [3.0,3.8]\,meV for constant $R$ while it is [3.0,3.2]\,meV for the density-dependent $R$. In
general, the actual stable droplet configuration has to be determined by experiments. Since
the density-dependent droplet radius is consistent with the properties of macroscopic
electron--hole droplets, we use equation~(\ref{eq:R_dens}) to study the properties of quantum
droplets.

Figure \ref{Fig8}(a) shows again the ground-state energy of the quantum droplet as function of
electron--hole density $\rho_{eh}$ for the density-dependent $R$. The dashed lines continue the
energy levels after the next higher quantum droplet state becomes the ground state. The
biexciton-binding energy is indicated by a horizontal line. We see that the binding energy of the
unstable $(n=3)$-liquid state remains smaller than the biexciton-binding energy even at elevated
$\rho_{eh}$ making it instable at all densities. In contrast to that, $E_{\mrm{bind}}$ of the
$(n=4)$- and $(n=5)$- liquid state is stronger than the biexciton value while it remains relatively
stable as the electron--hole density is increased.

\subsection{Ring structure of quantum droplets}

\begin{figure}[t]
\includegraphics*[scale=0.45]{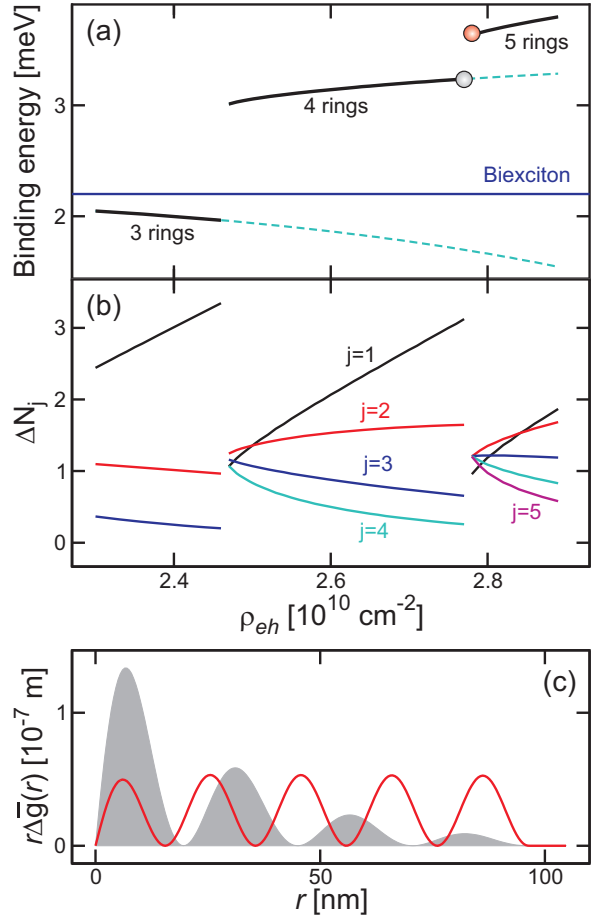}
\centering
\caption{Properties of quantum droplets.
(a) The ground-state energy (solid line) is presented as function of $\rho_{eh}$ for the
density-dependent $R$. The dashed lines denote excited states and the biexciton-binding energy is
marked by the horizontal line. (b) Number of correlated electron--hole pairs within the $j$-th ring
as function $\rho_{eh}$ from the first ($j=1$) up to the fifth ($j=5$) ring. (c) The
electron--hole pair-correlation function $r\Delta \bar{g}(r)$ is shown before (shaded
area) and after (solid line) the 4-to-5-ring droplet transition. These cases are indicated by
circles in frame (a).}
\label{Fig8}
\end{figure}

We also can analyze the number of correlated electron--hole pairs within the $j$-th ring of the
quantum droplet. Since
$S_{\mrm{drop}}\int \mrm{d}^2 r\, \Delta g(r)=S_{\mrm{drop}}\,2\pi\int\mrm{d} r\, r\Delta g(r)$
defines the total number of correlated pairs \cite{Kira:06b}, 
\begin{equation}
\label{eq:Nj_corr}
\Delta N_j=S_{\mrm{drop}}\,2\pi\int_{x_{j-1}}^{x_j} \mrm{d} r\, r\Delta g(r)
\end{equation}
is the number of correlated pairs within the $j$-th ring when $S_{\mrm{drop}}=\pi R^2$ is the area
of the quantum droplet. Figure \ref{Fig8}(b) shows $\Delta N_j$ as function of $\rho_{eh}$
from
the first up to the fifth ring. We see that the number of electron--hole pairs within the innermost
rings becomes larger, while it decreases within the outermost rings, as $\rho_{eh}$ is made larger.
Interestingly, each ring has roughly the same number of electron--hole pairs after the $n$-ring
droplet has become
the ground state via a sharp transition, compare with figure~\ref{Fig8}(a). More precisely,
$\Delta N_j$ is close to one such that the $n$-th quantum droplet state has about $n$ electron--hole
pairs
after the transition. Consequently, the $n$-ring quantum droplet has roughly $n$ electron--hole
pairs.
Therefore, already the first stable quantum droplet with $n=4$ rings has four correlated electrons
and
holes
showing that it is a highly correlated quasi-particle. As derived in \ref{Sec:Ndrop}, one can solve
analytically that for ring numbers up to $n=3$ the $n$-th quantum droplet state has very close $n$
correlated electron--hole pairs while the ratio $\Delta N/n$ converges towards 1.2 for a very large
ring number. 

Figure \ref{Fig8}(c) presents examples for the electron--hole pair-correlation
function $r \Delta \bar{g}(r)$ before (shaded area) and after (solid line) the 4-to-5-ring droplet
transition. The corresponding binding energies and electron--hole densities are indicated with
circles in figure~\ref{Fig8}(a). Before the transition, the oscillation amplitude of $r \Delta
\bar{g}(r)$ decreases as function of $r$ while after the transition the oscillation amplitude stays
almost constant indicating that the decay parameter $\kappa$ is close to zero, just after the
transition. This is consistent with our earlier observation that a $n$-ring quantum droplet emerges
only above a threshold density matching the density of the $\kappa=0$ state.

\section{Influence of electron--electron and hole--hole correlations}
\label{Sec:ee_corr}

So far, we have analyzed the properties of quantum droplets without electron--electron and
hole--hole correlations based on the assumption that electron--hole correlations dominate the
energetics. We will next show that this scenario is plausible also in dense interacting
electron--hole systems. We start by reorganizing the carrier--carrier correlations
$c^{\vec{q},\vec{k}',\vec{k}}_{\lambda,\lambda;\lambda,\lambda}$, defined in
equation~(\ref{eq:CE_corr}), into $\Delta \langle
a_{\lambda,\vec{K}+\vec{p}}^{\dagger}a_{\lambda,\vec{K}-\vec{p}}^{\dagger}a_{
\lambda,\vec{K}-\vec{p}'}a_{
\lambda,\vec{K}+\vec{p}'}\rangle$ using $\vec{k}=\vec{K}+\vec{p}$, $\vec{k}'=\vec{K}-\vec{p}$, and
$\vec{q}=\vec{p}-\vec{p}'$. In this form, we see that two annihilation (or creation)
operators assign a correlated carrier pair that has a center-of-mass momentum of $2\hbar\,\vec{K}$.
Like for electron--hole correlations, we concentrate on the case where the center-of-mass momentum
of the correlated pairs vanishes 
\begin{equation}
\label{eq:deltag_ee2} \fl 
 \Delta \langle
a_{\lambda,\vec{K}+\vec{p}}^{\dagger}a_{\lambda,\vec{K}-\vec{p}}^{\dagger}a_{
\lambda,\vec{K}-\vec{p}'}a_{
\lambda,\vec{K}+\vec{p}'}\rangle\equiv-\delta_{\vec{K},0}\,F^{\lambda}_{\vec{p
}  ,  \vec { p } ' } \qquad \Leftrightarrow \qquad
c^{\vec{q},\vec{k}',\vec{k}}_{\lambda,\lambda;\lambda,\lambda}=-\delta_{\vec{k}',-\vec{k}}
F^{\lambda}_{\vec{k},\vec{k}-\vec{q}}\,,
\end{equation}
that follows from a straight forward substitution
$\vec{K}=\textstyle{\frac{1}{2}}(\vec{k}+\vec{k}')$,
$\vec{p}=\textstyle{\frac{1}{2}}(\vec{k}-\vec{k}')$, and
$\vec{p}'=\textstyle{\frac{1}{2}}(\vec{k}-\vec{k}')-\vec{q}$. Since the
transformations $\vec{p}\rightarrow -\vec{p}$ and $\vec{p}'\rightarrow -\vec{p}'$ correspond to
exchanging creation and annihilation operators in $c_{\lambda,\lambda;\lambda,\lambda}$,
respectively, the $F^{\lambda}_{\vec{p},\vec{p}'}$ function must change its sign
with these transformations due to the Fermionic antisymmetry. In other words,
$F^{\lambda}_{\vec{p},\vec{p}'}$ must satisfy
\begin{equation}
\label{eq:F_cond} 
F^{\lambda}_{-\vec{p},\vec{p}'}=F^{\lambda}_{\vec{p},-\vec{p}'}=-F^{\lambda}_{\vec{p},\vec{p}'}=-F^
{\lambda}_ {-\vec{p}, -\vec{p}'}\,,
\end{equation}
when the sign of the momentum is changed.

Like for electron--hole correlations, carrier--carrier effects can be described through the
corresponding pair-correlation function
\begin{eqnarray}
&g_{\lambda}(\vec{r})\equiv\langle\Psi_{\lambda}^{\dagger}(\vec{r})\Psi_{\lambda}^{\dagger}
(0)\Psi_ {
\lambda}(0)\Psi_{\lambda}(\vec{r})\rangle=\rho_{\lambda}^2-f^2_{\lambda}(\vec{r})+\Delta
g_{\lambda}(\vec{r})\,, \qquad \\
&f_{\lambda}(\vec{r})\equiv\frac{1}{S}\sum_{\vec{k}}f^{\lambda}_{\vec{k}}\,\e^{-\mrm{i}\vec{k}
\cdot\vec {
r}
}\,, \quad \mrm{with}\quad \lambda=e, h\,,
\end{eqnarray}
where we have applied homogeneous conditions, used the definition (\ref{eq:field}), and introduced
$f_{\lambda}(\vec{r})$ as the Fourier transformation of $f_{\vec{k}}^{\lambda}$. The
first term describes again a plasma contribution analogously to the first part in the
electron--hole pair-correlation function (\ref{eq:pair_corr_fct}). The correlated contribution is
defined by
\begin{eqnarray}
\label{eq:delta_gee} 
\Delta
g_{\lambda}(\vec{r})\equiv\frac{1}{S^2}\sum_{\vec{K},\vec{p},\vec{p}'}\Delta\ev{a^{\dagger}_{
\lambda , \vec {K}+\vec{p}}a^{\dagger}_{\lambda,\vec
{K}-\vec{p}}a_{\lambda,\vec{K}-\vec{p}'}a_{\lambda,\vec{K}+\vec{p}'}}\,\e^{\mrm{i}(\vec{p}-\vec{p}
')\cdot\vec{r}} \nonumber \\ \qquad\quad\, =-\frac{1}{S^2}\sum_{\vec{p}, \vec{p}'}
F^{\lambda}_{\vec{p},\vec{p}'}\,\e^{\mrm{i}(\vec{p}-\vec{p}')\cdot\vec{r}}\,,
\end{eqnarray}
where we have applied the condition (\ref{eq:deltag_ee2}). We note that $\Delta g_{\lambda}(\vec{r})$
vanishes at $\vec{r}=0$ due to the Pauli-exclusion principle among Fermions, enforced by
equation~(\ref{eq:F_cond}).

Due to the
conservation law (\ref{eq:cons2}), the electron and hole distributions $f^{e}_{\vec{k}}$ and $f^{h}_{\vec{k}}$ become
different only when the electron--electron and hole--hole correlations are different. To study how the carrier--carrier correlations modify the
overall energetics, we assume identical electron--electron and hole--hole correlations
$F^{e}_{\vec{p},\vec{p}'}=F^h_{\vec{p},\vec{p}'}$ to simplify the book-keeping. With this
choice, equations~(\ref{eq:cons2}) and (\ref{eq:deltag_ee2}) imply identical distributions that
satisfy
\begin{equation}
\label{eq:cons_ee} 
\left(f_{\vec{k}}-\textstyle{\frac{1}{2}}\right)^2+g_{\vec{k},\vec{k}}+F_{\vec{k},\vec{k}}
=\textstyle{\frac { 1 } {
4}}\,, \qquad F_{\vec{k},\vec{k}}\equiv F^{e}_{\vec{k},\vec{k}}=F^{h}_{\vec{k},\vec{k}}\,.
\end{equation}
We see that also carrier--carrier correlations modify $f_{\vec{k}}$ via a diagonal
$F_{\vec{k},\vec{k}}$, just like $g_{\vec{k},\vec{k}}$.

In the same way, the generalized Wannier equation (\ref{eq:gen_Wannier}) is modified through
the presence of carrier--carrier correlations in the form of equation~(\ref{eq:F_cond}). By
inserting equations~(\ref{eq:deltag_ee2}) and (\ref{eq:cons_ee}) into
equation~(\ref{eq:gen_Wannier}), the
original $E_{\vec{k}}$ and $V^{\mrm{eff}}_{\vec{k},\vec{k}'}$ can simply be replaced by
\begin{equation}
\label{eq:E_renorm_ee}
 E_{\vec{k}}\equiv\left[\frac{\hbar^2\vec{k}^2}{2\mu}-2\sum_{\vec{k}'}V_{\vec{k}
-\vec{k}'}f_{\vec{k}'}\right]\left(1-2f_{\vec
{ k }
}\right)+2\sum_{\vec{k}'}V_{\vec{k}-\vec{k}'}\,\left(g_{\vec{k},\vec{k}'}+F_{\vec{k},\vec{k}'}
\right)\,,
\end{equation}
\begin{equation}
\label{eq:VS_ee}
V^{\mrm{eff}}_{\vec{k},\vec{k}'}\equiv
\left(1-2f_{\vec{k}}\right)
V_{\vec{k}-\vec{k}'}\left(1-2f_{
\vec{k}'}\right)+2\left(g_{\vec{k},\vec{k}'}+F_{\vec{k},\vec{k}'}\right)V_{\vec{k}
-\vec{k}'}\,,
\end{equation}
to fully account for the carrier--carrier contributions.

As a general property, the repulsive Coulomb interaction tends to extend the $\vec{r}$-range where
the presence of multiple carriers is Pauli blocked. In other words, carrier--carrier correlations
build up to form a correlation hole to $g_{\lambda}(\vec{r})$. To describe this principle effect,
we use an ansatz
\begin{equation}
\label{eq:ee_ansatz} 
F_{\vec{k},\vec{k}'}\equiv
F_0^2\cos(\theta_\vec{k}-\theta_{\vec{k}'})\,\e^{-l_{\mrm{c}}(|\vec{k}|-|\vec{k}'|)}\,,
\end{equation}
that satisfies the antisymmetry relations (\ref{eq:F_cond}). The strength of the correlation is
determined by $F_0$ and $l_{\mrm{c}}$ corresponds to a correlation length. As
equation~(\ref{eq:ee_ansatz}) is inserted to equation~(\ref{eq:delta_gee}), a straight forward
integration
yields 
\begin{equation}
\label{eq:ee_pair} 
\Delta g_{\lambda}(r)=-\frac{F_0^2}{(2\pi)^2}\frac{r^2}{(l_{\mrm{c}}^2+r^2)^3}\,,
\end{equation}
which is rotational symmetric and vanishes at $\vec{r}=0$, as it should for homogeneous
Fermions.

To compute the quasi-particle energetics with carrier--carrier correlations, we
use the same quantum droplet state (\ref{eq:phi_drop}) as computed for vanishing carrier--carrier
correlations in section~\ref{Sec:Epair_drop}, i.e.~we keep the quantum droplet radius $R$, ring
number
$n$, and decay parameter $\kappa$ unchanged. For a given combination $(F_0,l_{\mrm{c}})$, we then
adjust the strength of the electron--hole correlations $g_0$ such that
$g_{\vec{k},\vec{k}}+F_{\vec{k},\vec{k}}$ is maximized,
i.e.~$\max[g_{\vec{k},\vec{k}}+F_{\vec{k},\vec{k}}] =\textstyle{\frac{1}{4}}$, according to
equation~(\ref{eq:cons_ee}). In analogy to section~\ref{Sec:Epair_drop}, this yields a vanishing
$(1-2f_{\vec{k}})$ at one momentum state. Since $F_{\vec{k},\vec{k}}$ is positive, the presence of
carrier--carrier correlations must be compensated by reducing the magnitude of the electron--hole
correlations $g_{\vec{k},\vec{k}}$. Additionally, equation~(\ref{eq:cons_ee}) modifies the
electron--hole
distribution $f_{\vec{k}}$ and the electron--hole density in comparison to the case with vanishing
$F_{\vec{k},\vec{k}}$.

\begin{figure}[t]
\includegraphics*[scale=0.45]{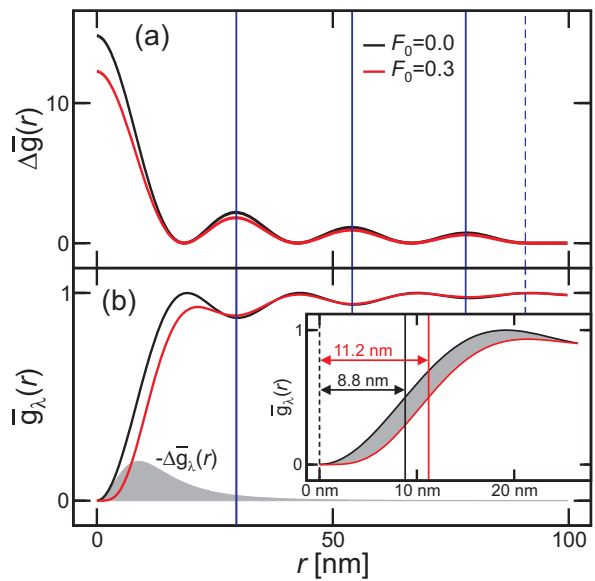}
\centering
\caption{Effect of carrier--carrier correlations on the quantum-droplet energetics.
(a) Normalized electron--hole pair-correlation function without ($F_0=0$, black line) and
with ($F_0=0.3$, $l_{\mrm{c}}=12.5$\,nm, red line) carrier--carrier correlations. The
quantum-droplet
state has $R=90.8$\, nm, $n=4$ rings, and $\rho_{eh}=2.5\times 10^{10}$\,nm ($\rho_{eh}=2.7\times
10^{10}$\,nm) for
$F_0=0$ ($F_0=0.3$). The maxima of $\Delta \bar{g}(r)$ are indicated by the vertical lines. (b) The
corresponding normalized carrier--carrier pair-correlation function $\bar{g}_{\lambda}(r)$. The pure
correlated contribution $-\Delta \bar{g}_{\lambda}(r)$ for $F_0=0.3$ is shown as a shaded area.
Inset: Same data as in (b) up to the first Friedel
oscillation $r=26.6$\,nm together with the half-widths.}
\label{Fig9}
\end{figure}

Figure \ref{Fig9}(a) shows the normalized electron--hole pair-correlation function $\Delta
\bar{g}(r)$ for vanishing carrier-carrier correlations ($F_0=0$, black line). The vertical
lines
indicate the maxima of $\Delta \bar{g}(r)$ identifying the centers of the liquid-state rings.
The
quantum droplet state has a radius of $R=90.8$\,nm, $n=4$ rings, and an electron--hole density of
$\rho_{eh}=2.5\times 10^{10}\,\mrm{cm}^{-2}$. The corresponding result for nonvanishing
carrier--carrier correlations with $F_0=0.3$ and $l_{\mrm{c}}=12.5$\,nm is plotted as red line. The
presence of carrier--carrier correlations increases the electron--hole density to
$\rho_{eh}=2.7\times 10^{10}\,\mrm{cm}^{-2}$ due to the normalization procedure described above.
We see that the presence of carrier--carrier correlations reduces the amplitude of the ring-state
oscillations in $\Delta \bar{g}(r)$ only slightly. This suggests that carrier--carrier
correlations play a minor role in the build up of electron--hole correlations in quantum droplets. 

The corresponding normalized carrier--carrier pair-correlation function
$\bar{g}_{\lambda}(r)\equiv g_{\lambda}(r)/\rho_{eh}^2$ is
presented in figure~\ref{Fig9}(b) without ($F_0=0$, black line) and with ($F_0=0.3$, red line)
carrier--carrier correlations. Additionally, the pure correlated contribution $-\Delta
\bar{g}_{\lambda}(r)\equiv -\Delta g_{\lambda}(r)/\rho_{eh}^2$ for $F_0=0.3$ is shown as a shaded
area. Even without carrier--carrier
correlations, $\bar{g}_{\lambda}(r)$ shows a range of Pauli blocked carriers at short
distances
followed by the Friedel oscillations \cite{Friedel:56}. Interestingly, $\bar{g}_{\lambda}(r)$
dips
at exactly the same positions where $\Delta \bar{g}(r)$ peaks indicated by vertical lines in
figure~\ref{Fig9}. Consequently, the carriers try to avoid each other within the rings of the
quantum droplets, which is clearly related to the Fermion character of electrons. We observe that
the
presence of $\Delta \bar{g}_{\lambda}(r)$ increases the range of Pauli-blocked carriers. To
show the
range of Pauli blocking, the inset of figure~\ref{Fig9}(b) plots the same data up to the first
Friedel oscillation $r=26.6$\,nm. To quantify Pauli blocking, we determine the half-width value
where $g_{\lambda}(r_{1/2})=\textstyle{\frac{1}{2}}\rho^2_{eh}$. We find that $r_{1/2}$ increases
from $8.8$\,nm for $F_0=0$ to $11.2$\,nm for $F_0=0.3$, i.e. the correlation hole increases the
range of Pauli blocking by roughly $27\,\%$ which is significant.

In the next step, we compute the ground-state energy of pair excitations from the generalized Wannier equation
(\ref{eq:gen_Wannier}) with the $f_{\vec{k}}$, $g_{\vec{k},\vec{k}'}$, and $F_{\vec{k},\vec{k}'}$
entries (\ref{eq:E_renorm_ee})--(\ref{eq:VS_ee}). The actual energy per excited particle follows
from equation~(\ref{eq:Epp_lambda}) and this is compared against the exciton binding deduced as in
section~\ref{Sec:X}. The results produce a quantum droplet energy that grows from $2.99$\,meV to
$3.08$\,meV
as the carrier--carrier correlations are included. The small increase shows
that the correlated arrangement of the carriers saves energy. However, carrier--carrier correlations
change the quantum droplet binding only by $3.3\,\%$, for the studied case. In other words, even a
large
correlation hole $\Delta g_{\lambda}(\vec{r})$ cannot affect much the energetics of the quantum
droplet,
which justifies the assumption of neglecting carrier--carrier correlations for quantum droplets.

\section{Discussion}

We have developed a systematic method to
compute the pair-excitation energetics of many-body states based on the
correlation-function formulation of quasi-particles. In particular, we
have generalized the Wannier equation to compute the energy per excited electron--hole pair of a
many-body state probed by a weak pair excitation of a quasi-particle. As an unconventional aspect,
we determine the many-body state via the pair-correlation function $g(\vec{r})$ and work out the
lower-order expectation values self-consistently, based on $g(\vec{r})$, not the other way around.
As a major benefit, $g(\vec{r})$ characterizes the many-body state and its energetics, which allows
us to identify the properties of different quasi-particles directly.

We have applied the scheme to
study especially the energetics and properties of quantum droplets as a new quasi-particle. Our
computations show that the pair-excitation energetics of quantum droplets has discrete bands that
appear
as sharp transitions. Additionally, each ring contains roughly
one electron--hole pair and only quantum droplets with more than 4 rings, i.e., electron--hole pairs
are
stable. We also show that the energy structure of quantum droplets originates dominantly from
electron--hole correlations because the carrier--carrier correlations increase the exciton
energy only slightly.

The developed method can be used more generally to determine the characteristic quasi-particle
energies based on the correlation function. As further examples, we successfully analyze the
energetics of the degenerate Fermi gas and high-density excitons. We also have extended the method
to analyze coherent quasi-particles. As possible new directions, one can study different
pair-excitation schemes to analyze the role of,
e.g., spin. In this connection, one expects to detect bonding and antibonding branches for
quasi-particles such as biexcitons. In general, the approach is limited only by the user's knowledge
of the pair-correlation function. It also might be interesting to develop the approach to the
direction where quasi-particles are identified via $N$-particle correlations to systematically
analyze how the details of highly correlated states affect the excitation energetics and the
response in general.

\ack
M.~K. acknowledges support from the Deutsche Forschungsgemeinschaft.

\begin{appendix}

\section{Connection of correlations and expectation values}

\label{sec:cons} 

We first analyze a normally ordered $(N+1)$-particle expectation value
\begin{equation}
\label{eq:N_exp1}
\ev{\hat{N}+1}\equiv\ev{a^{\dagger}_{\lambda_1,\vec{k}_1}\dots
a^{\dagger}_{\lambda_N,\vec{k} _N}\,\hat{N}_{\mrm{tot}}\,a^{\dagger}_{\lambda'_N,\vec{k}'_N}\dots
a^{\dagger}_{\lambda'_1,\vec{k}'_1}}\,,
\end{equation}
that contains the total number operator
$\hat{N}_{\mrm{tot}}\equiv\sum_{\vec{k},\lambda}a^{\dagger}_{\lambda,\vec{k}}a_{\lambda,\vec{k}}$.
Since $\hat{N}_{\mrm{tot}}$ contains \textit{all} electronic states, it produces
\begin{equation}
\label{eq:a2}
\hat{N}_{\mrm{tot}}\hat{\rho}_{\mathcal{N}}=\mathcal{N}\hat{\rho}_{\mathcal{N}}
\end{equation}
for all states $\hat{\rho}_{\mathcal{N}}$ containing $\mathcal{N}$ carriers within all bands of the
system. Since we may consider only cases where the total number of carriers is conserved, we may
limit the analysis to the states $\hat{\rho}_{\mathcal{N}}$ from here on.

By applying the commutator relation
$\textstyle{\left[\hat{N}_{\mrm{tot}},a_{\lambda,\vec{k}}\right]_{-}=-a_{\lambda,\vec{k}}}$  $N$
times, equation~(\ref{eq:N_exp1}) becomes
\begin{equation}
\label{eq:N_1} \fl
\ev{\hat{N}+1}=-N\ev{\hat{O}_N}+\ev{\hat{O}_N\hat{N}_{\mrm{tot}}}\,,
\quad \mrm{with} \quad \hat{O}_N\equiv a^{\dagger}_{\lambda_1,\vec{k}_1}\dots
a^{\dagger}_{\lambda_N,\vec{k} _N}a^{\dagger}_{\lambda'_N,\vec{k}'_N}\dots
a^{\dagger}_{\lambda'_1,\vec{k}'_1}\,.
\end{equation}
Using the property (\ref{eq:a2}), we find
\begin{equation}
\label{eq:RHS}
\ev{\hat{O}_N\hat{N}_{\mrm{tot}}}=\mrm{Tr}[\hat{O}_N\hat{N
} _
{\mrm{tot}} \hat { \rho } _ {
\mathcal{N}}]=\mrm{Tr}[\hat{O}\mathcal{N}\hat{\rho}_{\mathcal{N}}]=\mathcal{N}\ev{\hat{O}_N}\,.
\end{equation}
By combining the result (\ref{eq:RHS}) with (\ref{eq:N_exp1}) and (\ref{eq:N_1}), we obtain a
general reduction formula \cite{Hoyer:04} 
\begin{equation}
\label{eq:cons_gen} 
\sum_{\vec{k}',\lambda'}\ev{a^{\dagger}_{\lambda_1,\vec{k}_1}\dots
a^{\dagger}_{\lambda_N,\vec{k}_N}
a^{\dagger}_{\lambda',\vec{k}'}a_{\lambda',\vec{k}'}a_{\lambda_N',\vec{k}_N'}\dots
a_{\lambda'_1,\vec{k}'_1}}=(\mathcal{N}-N)\ev{\hat{\mathcal{O}}}\,,
\end{equation}
that directly connects $N$ and $(N+1)$-particle expectation values.

For $N=1$, equation~(\ref{eq:cons_gen}) becomes
\begin{equation}
\label{eq:gen_con}
\sum_{\vec{k}',\lambda'}\ev{a^{\dagger}_{\lambda,\vec{k}}a^{\dagger}_{\lambda',\vec{k}'}a_{\lambda',
\vec{k}'}a_{\lambda,\vec{k}}}=(\mathcal{N}-1)\ev{a^{\dagger}_{\lambda,\vec{k}}a_{\lambda,\vec{k}}}\,
.
\end{equation}
We then express the two-particle contribution exactly in terms of the Hartree--Fock factorization
\cite{Book:11} and the two-particle correlations (\ref{eq:CE_corr}) and assume homogeneous
conditions where all coherences vanish. By using a two-band model,
equation~(\ref{eq:gen_con}) yields then
\begin{eqnarray}
\label{eq:cons} 
 &\left(f^{e}_{\vec{k}}-\textstyle{\frac{1}{2}}\right)^2+\sum_{\vec{k}'}
\left(c^ { \vec { k } -\vec {
k } ' , \vec{k}',\vec{k}}_{eh}-c^{0,\vec{k}',\vec{k}}_{c,c;c,c}\right)= \textstyle{\frac { 1
} { 4 }} \,, \qquad \nonumber \\
 &\left(f^{h}_{\vec{k}}-\textstyle{\frac{1}{2}}\right)^2+\sum_{\vec{k}'}
\left(c^{\vec{k}'-\vec
{ k
}, \vec{k},\vec{k}'}_{eh}-c^{0,\vec{k}',\vec{k}}_{v,v;v,v}\right)= \textstyle{\frac { 1 }
{ 4 }}  \,,
\end{eqnarray}
for electrons $(\lambda=c)$ and holes $(\lambda=v)$, respectively. With the help
of equation~(\ref{eq:eq_com}), equation~(\ref{eq:cons}) casts into the form 
\begin{equation}
\label{eq:cons1} \fl
 \left(f^{e}_{\vec{k}}-\textstyle{\frac{1}{2}}\right)^2+g_{\vec{k},\vec{k}}-\sum_{\vec{k}'}c^{
0 , \vec { k } ' , \vec { k } } _ {c,c;c,c}= \textstyle{\frac { 1
} { 4 }} \,, \qquad
 \left(f^{h}_{\vec{k}}-\textstyle{\frac{1}{2}}\right)^2+g_{\vec{k},\vec{k}}-\sum_{\vec{k}'}c^{
0 , \vec { k } ' , \vec { k } } _ {v,v;v,v}= \frac { 1 }
{ 4 }  \,,
\end{equation}
that connects the density distributions with the pair-wise correlations.

\section{\label{app1} Probe-induced quantities}

To compute the probe-induced electron--hole density and polarization, we use the following general
properties of the displacement operator (\ref{eq:disp}) \cite{Book:11,Kira:06b}
\begin{eqnarray}
\label{eq:D_props}
&D^{\dagger}[\psi]a_{v,\vec{k}}D[\psi]=\cos(\varepsilon\,|\psi_{\vec{k}}|)a_{v,\vec{k}}
-\e^{-\mrm{i}\varphi_{\vec{k}}}\sin(\varepsilon\,|\psi_{\vec{k}}|)a_{c,\vec{k}}\,, \nonumber
\\
&D^{\dagger}[\psi]a_{c,\vec{k}}D[\psi]=\cos(\varepsilon\,|\psi_{\vec{k}}|)a_{c,\vec{k}}
+\e^ { \mrm { i }
\varphi_ {\vec{k}}}\sin(\varepsilon\,|\psi_{\vec{k}}|)a_{v,\vec{k}}\,.
\end{eqnarray}
Transformation (\ref{eq:D_props}) allows us to construct the density- and polarization-induced pair
excitations exactly.
More specifically, we start from the expectation value
\begin{equation}
\label{eq:D_exp} \fl
\ev{a^{\dagger}_{\lambda,\vec{k}}a_{\lambda',\vec{k}}}_{\psi}\equiv\mrm{Tr}\left[a^{\dagger}_{
\lambda ,
\vec{k}}a_{\lambda',\vec{k}}\hat{D}[\psi]\hat{\rho}_{\mrm{MB}}\hat{D}^{\dagger}[\psi]\right]
=\mrm{Tr}\left[\hat{D}^{\dagger}[\psi]a^{\dagger}_{\lambda,\vec{k}}\hat{D}[\psi]\hat{D}^{\dagger}
[ \psi]a_{\lambda',\vec{k}}\hat{D}[\psi]\hat{\rho}_{\mrm{MB}}\right]\,,
\end{equation}
where we have utilized cyclic permutations under the trace and the unitary of the
displacement operator (\ref{eq:disp}). 

To compute the pair-excitation energy, we have to compute how all those single-particle
expectation values and two-particle correlations that appear in equation~(\ref{eq:Emb}) are modified
by the pair excitation. By inserting transformation (\ref{eq:D_props}) into
equation~(\ref{eq:D_exp}), we
can express any modified single-particle expectation value in terms of $\varepsilon$,
$\psi_{\vec{k}}$, and $f_{\vec{k}}$. The change in density and polarization becomes then
\begin{eqnarray}
\label{eq:probe_fp} 
&f_{\vec{k},\psi}\equiv\ev{a^{\dagger}_{c,\vec{k}}a_{c,\vec{k}}}_{\psi}-f^{e}_{
\vec { k } } =\ev {
a_{v,\vec{k}}a^{\dagger}_{v,\vec{k}}}_{\psi}-f^{h}_{\vec{k}}
=\sin^2(\varepsilon\,|\psi_{\vec{k}}|)\left(1-f^{e}_{\vec{k}}-f^{h}_{\vec{k}}\right)\,,
\nonumber \\
&P_{\vec{k},\psi}\equiv\ev{a^{\dagger}_{v,\vec{k}}a_{c,\vec{k}}}_{\psi}=\e^{\mrm{i}
\varphi_{\vec{k}}}\sin(\varepsilon\, |\psi_ { \vec { k } }|)
\cos(\varepsilon\,|\psi_{\vec{k}}|)\left(1-f^{e}_{\vec{k}}-f^{h}_{\vec{k}}\right)\,,
\end{eqnarray}
respectively. Since the many-body state $\hat{\rho}_{\mrm{MB}}$ is probed by a weak laser
pulse, we apply the weak-excitation limit $\varepsilon\ll 1$, producing
\begin{equation}
\label{eq:probe_p&f} \fl
  P_{\vec{k},\psi}=
\left(1-f^{e}_{\vec{k}}-f^{h}_{\vec{k}}\right)\varepsilon\,\psi_{\vec{k}}+\mathcal{O}
(\varepsilon^3)\,, \quad
f_{\vec{k},\psi}=\left(1-f^{e}_{\vec{k}}-f^{h}_{\vec{k}}\right)\varepsilon^2\,
|\psi_{\vec{k}}|^2+\mathcal{O}(\varepsilon^3)\,,
\end{equation}
to the leading order.

Following the same derivation steps as above, we find that the pair excitations change the
electron--hole correlation by
\begin{eqnarray}
\label{eq:probe_corr} \fl
 c^{\vec{q},\vec{k}',\vec{k}}_{eh,\psi}\equiv
\Delta\ev{a^{\dagger}_{c,\vec{k}}a^{\dagger}_{v,\vec{k}'}a_{c,\vec{k}'+\vec{q}}a_{v,\vec{k}-\vec{q}}
}_{\psi}-c^{\vec{q},\vec{k}',\vec{k}}_{eh} \nonumber \\
 \fl\qquad\,\,\, =\varepsilon\left[c^{-\vec{q},\vec{k},\vec{k}'}_{v,v;v,c}
\psi_{\vec{k}}^{\star}+\left(c^{-\vec{q},\vec{k}'-\vec{q},\vec{k}+\vec{q}}_{v,v;v,c}\right)^{\star}
\psi_{\vec{k}'+\vec{q}}-c^{-\vec{q},\vec{k},\vec{k}'}_{v,c;c,c}\psi_{\vec{k}-\vec{q}}^{\star}
-\left(c^{-\vec{q},\vec{k}'-\vec{q},\vec{k}+\vec{q}}_{v,c;c,c}\right)^{\star}\psi_{\vec{k}'}\right
]\nonumber \\
\fl\qquad\,\,\,\,
+\,\varepsilon^2\left[c^{-\vec{q}+\vec{k}-\vec{k}',\vec{k}',\vec{k}}_{eh}\psi_{\vec{k}'+\vec{q}}
\psi_ {
\vec{k}-\vec{q}}^{\star}+\left(c^{\vec{q}-\vec{k}+\vec{k}',\vec{k}'-\vec{q},\vec{k}+\vec{q}}_{eh}
\right)^{\star}\psi_{\vec{k}'}\psi_{\vec{k}}^{\star}\right. \nonumber \\
\fl\qquad\,\,\,\, \left.
\qquad-\textstyle{\frac{1}{2}}c^{\vec{q},\vec{k}',\vec{k}}_{eh}\left(|\psi_{\vec{k}}|^2+|\psi_{\vec{
k } '} |^2+|\psi_ { \vec{k}-\vec{q}}|^2+|\psi_{\vec{k}'+\vec{q}}|^2\right)\right. \nonumber
\\ 
\fl\qquad\,\,\,\, \left.\qquad +
c^{\vec{q},\vec{k}',\vec{k}}_{c,c;c,c}\psi_{\vec{k}'}\psi_{\vec{k}-\vec{q}}^{\star}+c^{\vec{q},\vec{
k}',\vec{k}}_{v,v;v,v}\psi_{\vec{k}'+\vec{q}}\psi_{\vec{k}}^{\star}\right. \nonumber
\\ 
\fl\qquad\,\,\,\, \left.\qquad-c^{\vec{q},\vec{k}',\vec{k}}_{v,
v;c,c}\psi_{\vec{k}}^{\star}\psi_{\vec{k}-\vec{q}}^{\star}
-\left(c^{-\vec{q},\vec{k}'-\vec{q},\vec{k
}+\vec{q}}_{v,v;c,c}\right)^{\star}\psi_{\vec{k}'}\psi_{\vec{k}'+\vec{q}}
\right]+\mathcal{O}(\varepsilon^3)
\end{eqnarray}
out of the initial many-body correlation $c^{\vec{q},\vec{k}',\vec{k}}_{eh}$. Besides the
correlations (\ref{eq:CE_corr}), equation~(\ref{eq:probe_corr}) contains also coherent
two-particle correlations:
\begin{eqnarray}
\label{eq:coh_corr}
\fl
 c_{v,c;c,c}^{\vec{q},\vec{k}',\vec{k}}\equiv\Delta\ev{a^{\dagger}_{v,\vec{k}}a^{\dagger}_{c,\vec{k
}'}a_{c,\vec{k}'+\vec{q}}a_{c,\vec{k}-\vec{q}}}\,, \quad
 c_{v,v;v,c}^{\vec{q},\vec{k}',\vec{k}}\equiv\Delta\ev{a^{\dagger}_{v,\vec{k}}a^{\dagger}_{v,\vec{k
}'}a_{v,\vec{k}'+\vec{q}}a_{c,\vec{k}-\vec{q}}}\,, \nonumber \\
\fl
c_{v,v;c,c}^{\vec{q},\vec{k}',\vec{k}}\equiv\Delta\ev{a^{\dagger}_{v,\vec{k}}a^{\dagger}_{v,\vec{k
} ' } a_ {
c,\vec{k}'+\vec{q}}a_{c,\vec{k}-\vec{q}}}\,.
\end{eqnarray}
From these, $c_{v,c;c,c}^{\vec{q},\vec{k}',\vec{k}}$ and $c_{v,v;v,c}^{\vec{q},\vec{k}',\vec{k}}$
describe correlations between polarization and density while
$c_{v,v;c,c}^{\vec{q},\vec{k}',\vec{k}}$ corresponds to the coherent biexciton
amplitude. Therefore, also the coherent two-particle correlations
(\ref{eq:coh_corr}) contribute to the pair-excitation spectroscopy even though they do not
influence the initial many-body energy (\ref{eq:Emb}). The remaining
$c^{\vec{q},\vec{k}',\vec{k}}_{c,c;c,c}$ and $c^{\vec{q},\vec{k}',\vec{k}}_{v,v;v,v}$ transform
analogously. With the help of equations~(\ref{eq:probe_p&f})--(\ref{eq:probe_corr}) we can then
construct
exactly the energy change (\ref{eq:Epro_ex}) induced by the pair-wise excitations.

\section{Generalized Wannier equation with coherences}

\label{Sec:Wannier_com} 

As the exact relations (\ref{eq:probe_p&f})--(\ref{eq:probe_corr}) are inserted to the system
energy (\ref{eq:Epro_ex}), we obtain the pair-excitation energy exactly
\begin{eqnarray}
 \label{eq:Epro_ex2} \fl
 E_{\mrm{pro}}[\psi]=E_{\mrm{pro}}^{\mrm{coh}}[\psi]+E_{\mrm{pro}}^{\mrm{inc}}[\psi]+\mathcal{O}
(\varepsilon^3)\,, \nonumber
\\
\fl
E^{\mrm{coh}}_{\mrm{pro}}\equiv
2\varepsilon\sum_{\vec{k}}\left[\tilde{E}_{\vec{k}}\re[P_{\vec{k}}\psi^{\star}_{\vec{k}}]
-\sum_ {\vec { k } ' } V_ {
\vec{k}-\vec{k}'}\left(1-f^e_{\vec{k}}-f^h_{\vec{k}}\right)\re[P_{\vec{k}'}\psi^{\star}_{\vec{k}}]
+ \re[\Gamma_{\vec{k}}\psi_{\vec{k}}^{\star}]\right] \nonumber \\
\fl
\qquad \, -\,
2\varepsilon^2\sum_{\vec{k},\vec{k}'}V_{\vec{k}-\vec{k}'}\left(\re[P_{\vec{k}}P_{\vec{k}'}
\psi^ { \star}_
{\vec{k}}(\psi^{\star}_{\vec{k}'}-\psi^{\star}_{\vec{k}})]-\re[P_{\vec{k}}P_{
\vec{k}'}^{\star}]|\psi_{\vec{k}}|^2+\re[P_{\vec{k}'}P_{\vec{k}}^{\star}\psi_{\vec{k}}\psi_{
\vec{k}'}^{\star}]\right) \nonumber \\
\fl
\qquad \,\, +\, \varepsilon^2\sum_{\vec{k},\vec{k}',\vec{q}}V_{\vec{q}}\,
\re[\left(c^{\vec{q},\vec{k}'-\vec{q},\vec{k}+\vec{q}}_{v,v;c,c}+c^{\vec{q},\vec{k}',\vec{k}}_{v,v;c
,c}-2
c^{\vec{q},\vec{k}'-\vec{q},\vec{k}}_{v,v;c,c}\right)\psi^{\star}_{\vec{k}}\psi^{\star}_{\vec{k}'}]
\,, \nonumber
\\
\fl
E^{\mrm{inc}}_{\mrm{pro}}\equiv
\varepsilon^2\sum_{\vec{k}}\bar{E}_{\vec{k}}|\psi_{\vec{k}}|^2-\varepsilon^2\sum_{\vec{k},\vec{k}'}
\bar{V}^{\mrm{ eff}}_{\vec{k
},\vec{k}'}\psi_{\vec{k}}\psi_{\vec{k}'}^{\star}\,, \nonumber
\\
\fl
\qquad \,\,\, +\,\varepsilon^2\sum_{\vec{k},\vec{k}',\vec{q}}V_{\vec
{q}}\left(c^{
\vec{q},\vec{k}',\vec{k}}_{c,c;c,c}\psi_{\vec{k}}\psi_{\vec{k}-\vec{q}}^{\star}+c^{\vec{q},\vec{k}',
\vec{k}}_{v,v;v,v}\psi_{\vec{k}-\vec{q}}\psi_{\vec{k}}^{\star}\right)\,,
\end{eqnarray}
where we have divided $E_{\mrm{pro}}[\psi]$ into coherent (coh) and incoherent (inc)
contributions. The coherent contribution $E^{\mrm{coh}}_{\mrm{pro}}[\psi]$ includes
\begin{eqnarray}
 \Gamma_{\vec{k}}\equiv\sum_{\vec{k}',\vec{q},\nu}V_{\vec{q}}\left[c^{\vec{q},\vec{k}',\vec{k}}_{v,
\nu;\nu,c}-\left(c^{\vec{q},\vec{k}',\vec{k}}_{c,
\nu;\nu,v}\right)^{\star}\right]\,,
\end{eqnarray}
that is exactly the same as the microscopically described Coulomb scattering term in the
semiconductor Bloch equations \cite{Kira:06b}. The incoherent part
$E^{\mrm{inc}}_{\mrm{pro}}[\psi]$ and the coherent energy contain different renormalized kinetic
energies
\begin{eqnarray}
\label{eq:renorm_gen}
\bar{E}_{\vec{k}}&\equiv\tilde{E}_{\vec{k}}\left(1-f^e_{\vec{k}}-f^h_{\vec{k}}\right)+\sum_{\vec{k}'
, \vec{q}}V_{\vec{q}}\,\re[c^{\vec{q},\vec{k}',\vec{k}}_{c,c;c,c}+c^{
\vec{q},\vec{k}',\vec{k}}_{v,v;v,v}]
\nonumber \\
&+\sum_{\vec{k}',\vec{q}}V_{\vec{k}'+\vec{q}-\vec{k}}\left(\re[c^
{\vec{q},\vec{k}',\vec{k}}_{eh}]+\re[c^{-\vec{q},\vec{k},\vec{k}'}_{eh}]\right)\,, \nonumber \\
\tilde{E}_{\vec{k}}&=\frac{\hbar^2\vec{k}^2}{2\mu}-\sum_{\vec{k}'}V_{\vec{k}-\vec{k}'}\left(f^e_{
\vec {k}'}+f^h_{\vec{k}'}\right)\,,
\end{eqnarray}
respectively. We also have identified the effective Coulomb matrix element
\begin{eqnarray}
\label{eq:VS2} \fl
\bar{V}^{\mrm{eff}}_{\vec{k},\vec{k}'}\equiv\left(1-f^{e}_{\vec{k}}-f^{h}_{\vec{k}}\right)
V_{\vec{k}-\vec{k}'}\left(1-f^{e}_{
\vec{k}'}-f^{h}_{\vec{k}'}\right)-\sum_{\vec{k}',\vec{q}}V_{\vec{k}-\vec{k}'}\left(c^{\vec{q},\vec{k
}'-\vec{q},\vec{k}}_{eh}+c^{\vec{q},\vec{k}',\vec{k}+\vec{q}}_{eh}\right) \nonumber \\
\fl\qquad\,
-\sum_{\vec{k}',\vec{q}}V_{\vec{q}}\,\left(c^{\vec{q},\vec{k}'-\vec{q},\vec{k}}_{eh}+c^{\vec{q}    
\vec{k}',\vec{k}+\vec{q}}_{eh}-c^{\vec{q},\vec{k}'-\vec{q},\vec{k}+\vec{q}}_{eh}-c_{eh}^{\vec{
}\vec{  k} ',\vec{k}}\right), 
\end{eqnarray}
that contains the unscreened Coulomb interaction together with the phase-space filling
contribution $(1-f^{e}_{\vec{k}}-f^{h}_{\vec{k}})$ and electron--hole correlations
$c_{eh}^{\vec{q},\vec{k}',\vec{k}}$.

We then minimize the energy functional (\ref{eq:Epro_ex2}) as described in
section~\ref{Sec:approach}
to find a condition for the ground-state excitations. As a result, we obtain
\begin{eqnarray}
\label{eq:Wannier_gen2} \fl
 s_{\mrm{coh}}+\varepsilon E_{\mrm{coh}}[\psi]+\varepsilon E_{\mrm{inc}}[\psi]=\varepsilon
E_{\lambda}\psi_{\vec{k}}\,, \nonumber \\
\fl  s_{\mrm{coh}}\equiv \tilde{E}_{\vec{k}}
P_{\vec{k}}-\left(1-f^{e}_{\vec{k}}-f^{h}_{\vec{k}}\right)\sum_{\vec{k}'}V_{\vec{k}-\vec
{k}'}P_{
\vec{k}'}+\Gamma_{\vec{k}}\,, \nonumber \\
\fl E_{\mrm{coh}}[\psi]\equiv2\sum_{\vec{k}'}V_{\vec{k}-\vec{k}'}\left(P_{\vec{k}}P_{\vec{k}'}\psi^{
\star}_{\vec{k}}+\re[P_{\vec{k}}P_{\vec{k}'}^{\star}]\psi_{\vec{k}}\right)-2\sum_{\vec{k}'}V_{\vec{k
}-\vec{k}'}
\left(P_{\vec{k}}P_{\vec{k}'}\psi_{\vec{k}'}^{\star}+P_{\vec{k}}P_{\vec{k}'}^{\star}\psi_{\vec{k}'}
\right) \nonumber \\
\fl
\qquad\quad\,
+\,\sum_{\vec{k}',\vec{q}}V_{\vec{q}}\left(c^{\vec{q},\vec{k}',\vec{k}}_{v,v;c,c}-c^{\vec{q}, \vec {
k
} ' ,
\vec{k}+\vec{q}}_{v,v;c,c}\right)\left(\psi_{\vec{k}'}^{\star}-\psi_{\vec{k}'+\vec{q}}^{\star}
\right)\,,\nonumber \\
\fl
E_{\mrm{inc}}[\psi]\equiv\bar{E}_{\vec{k}}\psi_{\vec{k}}-\sum_{\vec{k}'}\bar{V}^{\mrm{eff}}_{\vec{
k},\vec{k}'}\psi_{\vec{k}'}+\sum_{\vec{k}',\vec{q}}V_{\vec{q}}\left(c^{\vec{q},\vec{k}',\vec{k}+\vec
{q}}_{c,c;c,c}\psi_{\vec{k}+\vec{q}}+c^{\vec{q},\vec{k}',\vec{k}}_{v,v;v,v}\psi_{\vec{k}-\vec{q}}
\right)\,.
\end{eqnarray}
We see that the presence of coherences generates the coherent source term $s_{\mrm{coh}}$ to
the generalized Wannier equation which is the dominant contribution in
equation~(\ref{eq:Wannier_gen2}). However, since $s_{\mrm{coh}}$ corresponds exactly to the
homogeneous part of the semiconductor Bloch equations \cite{Kira:06b}, it vanishes for stationary
$P_{\vec{k}}$. Therefore, the ground state of excitation must satisfy the generalized Wannier
equation 
\begin{eqnarray}
\label{eq:Wannier_gen3}
 E_{\mrm{coh}}[\psi]+E_{\mrm{inc}}[\psi]=E_{\lambda}\psi_{\vec{k}}\,.
\end{eqnarray}
In the main part, we analyze the pair excitations of incoherent many-body systems such that
$E_{\mrm{coh}}[\psi]$ is not present.

\section{\label{Sec:X_solver} Self-consistent exciton solver}

To find the wavefunction $\phi_{1s,\vec{k}}$ and the electron--hole distribution $f_{\vec{k}}$
that satisfy the ordinary density-dependent Wannier equation (\ref{eq:Wannier_1s}) and the
conservation law (\ref{eq:cons_final}), we define a gap equation as in
reference~\cite{Littlewood:96}
\begin{equation}
\label{eq:gap1} \fl
\Delta_{\vec{k}}\equiv \sum_{\vec{k}'}V_{\vec{k}-\vec{k}'}\phi_{1s,\vec{k}'}\,,\qquad
\epsilon_{\vec{k}}\equiv \frac{1}{2}\left(\tilde{E}_{\vec{k}}-E_{1s}\right)\,, \qquad
\Omega_{\vec{k}}=\sqrt{\epsilon_{\vec{k}}^2+\Delta_{\vec{k}}^2}\,.
\end{equation}
As a result, we obtain the integral equations
\begin{equation}
\label{eq:int_eq} 
P_{\vec{k}}=\frac{1}{2}\frac{\Delta_{\vec{k}}}{\Omega_{\vec{k}}}\,, \qquad
f_{\vec{k}}=\frac{1}{2}\left(1-\frac{\epsilon_{\vec{k}}}{\Omega_{\vec{k}}}\right)\,,
\end{equation}
which simultaneously satisfy the ordinary density-dependent Wannier equation (\ref{eq:Wannier_1s})
and the conservation law (\ref{eq:cons_final}). Equations (\ref{eq:gap1})--(\ref{eq:int_eq}) are
solved numerically by using the iteration steps
\begin{eqnarray}
\label{eq:it2} \fl
\Delta^{(n+1)}_{\vec{k}}=\sum_{\vec{k}'}V_{\vec{k}-\vec{k}'}P_{\vec{k}'}^{(n)}\,,\quad
\epsilon^{(n+1)}_{\vec{k}}=\frac{1}{2}\left(\frac{\hbar^2\vec{k}^2}{2\mu}-E_{1s}\right)\,, \quad
\Omega^{(n+1)}_{\vec{k}}=\sqrt{(\epsilon_{\vec{k}}^{(n+1)})^2+(\Delta_{\vec{k}}^{(n+1)})^2}\,,
\nonumber \\
P^{(n+1)}_{\vec{k}}=\frac{1}{2}\frac{\Delta^{(n+1)}_{\vec{k}}}{\Omega^{(n+1)}_{\vec{k}}}\,, \qquad
f^{(n+1)}_{\vec{k}}=\frac{1}{2}\left(1-\frac{\epsilon^{(n+1)}_{\vec{k}}}{\Omega^{(n+1)}_{\vec{k}}}
\right)\,.
\end{eqnarray}
One typically needs 40 iteration steps to reach convergence.

\section{ Number of correlated electron--hole pairs within droplet}

\label{Sec:Ndrop}

To compute the number of correlated pairs within the droplet close to the transition, we
start from the quantum droplet pair-correlation function defined by (\ref{eq:phi_drop}). Since the
decay
constant $\kappa$ is negligible small after each transition, see section~\ref{Subsec:G_drop}, we set
$\kappa=0$ in equation~(\ref{eq:phi_drop}), yielding
\begin{equation}
\label{eq:g_drop_S}
\phi(\vec{r})=J_0(x_n\textstyle{\frac{r}{n}})\,\theta(R-r)\,.
\end{equation}
The correlated electron--hole density is then given by \cite{Kira:06b}
\begin{equation}
\label{eq:corr_dens}
\Delta n\equiv\int\mrm{d}^2 r\,\Delta g(\vec{r})=2\pi
g_0^2\int_{0}^{R}\mrm{d}r\,r|J_0(x_n\textstyle{\frac{r}{R}})|^2=\pi g_0^2 R^2 [J_1(x_n)]^2\,,
\end{equation}
where we have introduced polar coordinates and used the properties of the Bessel
functions \cite{Arfken:12} in the last step.

To determine the parameter $g_0$ as function of the ring number $n$ and the droplet radius $R$, we
compute the Fourier transformation of $g_0 \phi(\vec{r})$, producing
\begin{equation}
\label{eq:FT_g0}
g_0 \phi_\vec{k}=g_0\int \mrm{d}^2 r\,
\phi(\vec{r})\,\e^{-\mrm{i}\vec{k}\cdot\vec{r}} =2 \pi
g_0\int_{0}^{R}\mrm{d}r\,r J_0(kr)J_0(x_n\textstyle{\frac{r}{R}})\,,
\end{equation}
where we have again introduced polar coordinates and identified
$J_0(kr)=2\pi\int_0^{2\pi}\mrm{d}\theta\,\e^{\mrm{i}kr\cos\theta}$ \cite{Arfken:12}. For a maximally
excited quantum droplet state, the maximum of $g_0 \phi_{\vec{k}}$ is $\max[g_0
\phi_{\vec{k}}]=\textstyle{\frac{1}{2}}$, based on the discussion in section~\ref{Sec:Epair_drop}. 
At the same time, the integral in equation~(\ref{eq:FT_g0}) is maximized for $k=x_n/R$. By
applying the orthogonality of Bessel functions, we obtain
\begin{equation}
\label{eq:FT_g02}
\max[g_0 \phi_\vec{k}]=\pi g_0 R^2[J_1(x_n)]^2=\textstyle{\frac{1}{2}}\,,
\end{equation}
such that $g_0$ can be written as
\begin{equation}
\label{eq:g0}
g_0=\left[2\pi R^2\,[J_1(x_n)]^2\right]^{-1}\,.
\end{equation}
By inserting equation~(\ref{eq:g0}) into equation~(\ref{eq:corr_dens}) and multiplication of $\Delta
n$
with the droplet area $S_{\mrm{drop}}\equiv\pi R^2$, the number of correlated pairs within
the droplet close to the transition becomes
\begin{equation}
\label{eq:corr_pairs}
\Delta N\equiv\pi R^2 \Delta n=\frac{1}{4[J_1(x_n)]^2}\,.
\end{equation}
This formula predicts that quantum droplets contain $\Delta N=3.4$, $\Delta N=4.6$, and $\Delta
N=5.9$ correlated electron--hole pairs for $n=3$, $n=4$, and $n=5$ rings, respectively. For ring
numbers larger than $n = 10$, $\Delta N$ approaches $1.2\,n$.

 \end{appendix}

\section*{References}

%

\end{document}